\documentclass[twocolumn]{aastex631}

\usepackage{bm}

\newcommand{\etaO}{\eta_\mathrm{O}}
\newcommand{\etaH}{\eta_\mathrm{H}}

\newcommand{\etaA}{\eta_\mathrm{A}}
\newcommand{\half}{{\frac{1}{2}}}

\newcommand{\LH}{L_\mathrm{H}}
\newcommand{\tcs}{\tilde{c}_\mathrm{s}}
\newcommand{\cA}{c_\mathrm{A}}
\newcommand{\machA}{\mathcal{M}_\mathrm{A}}

\begin{document}

\title{
Comparative Analysis of Hall Effect Implementations in Hall-Magnetohydrodynamics
}

\author[0000-0002-2707-7548]{Kazunari Iwasaki}
\affiliation{
Center for Computational Astrophysics, National Astronomical Observatory of Japan, Osawa, Mitaka, Tokyo 181-8588, Japan,
kazunari.iwasaki@nao.ac.jp
}
\author[0000-0001-8105-8113]{Kengo Tomida}
\affiliation{
Astronomical Institute, Tohoku University, Sendai, Miyagi, 980-8578, Japan
}




\begin{abstract}

There is no standard numerical implementation of the 
Hall effect, 
which is one of the non-ideal magnetohydrodynamic (MHD) effects. 
Numerical instability arises
when a simple implementation is used, 
in which the Hall electric field is added to 
the electric field to update magnetic fields without 
further modifications to the numerical scheme.
In this paper, several implementations proposed 
in the literature are compared to 
identify an approach that provides stable and accurate results.
We consider two types of implementations of the Hall effect.
One is a modified version of the Harten-Lax-van Leer 
method (Hall-HLL) in which 
the phase speeds of whistler waves are adopted as the signal speeds;
the other involves adding a fourth-order hyper-resistivity to a Hall-MHD code.
Based on an extensive series of test calculations,  
we found that hyper-resistivity yields more accurate results than the Hall-HLL,
particularly in problems where 
the whisler-wave timescale is shorter than the the timescale of physical processes of interest.
Through both von Neumann stability analysis and numerical experiments, 
an appropriate coefficient for 
the hyper-resistivity is determined.

\end{abstract}



\section{Introduction} \label{sec:intro}

The Hall effect is one of the non-ideal magnetohydrodynamic (MHD) effects.
It occurs in situations where 
electrons are coupled to magnetic fields, while 
ions are decoupled.
The characteristics of the Hall effect differ between 
fully ionized and weakly ionized gases \citep{PandeyWardle2008}.

An intriguing property of the Hall effect is that 
it modifies the dispersion relation of linear MHD waves, leading 
to generation of whistler and ion-cyclotron waves. 
In the long-wavelength limit, these waves transition into 
Alfv\'en waves. 
The phase speed of whistler waves increases proportionally to the wavenumber,
whereas that of ion-cyclotron waves remains constant at higher 
wavenumbers.

Numerically implementing the Hall effect presents significant 
challenges, unlike 
other non-ideal MHD effects 
(Ohmic resistivity and ambipolar diffusion), 
which are relatively simple to implement.
One reason is that the Hall effect causes waves to be dispersive.
In particular, as whistler waves with shorter wavelengths propagate faster,
grid-scale disturbances in magnetic fields can become significant under certain conditions.

A simple implementation of the Hall effect 
involves adding the Hall electric field to the 
electric field to update the magnetic field without further modifications to the numerical scheme.
\citet{Falle2003MNRAS.344.1210F} conducted a von Neumann stability analysis of such a naive
implementation and found it to be unconditionally unstable when 
using a first-order explicit time integrator (forward Euler integrator).

The stability of this simple implementation 
depends on the accuracy of the time integrator \citep{Kunz2013MNRAS.434.2295K}. 
While third-order time integrators are conditionally stable,
second-order time integrators may lead to numerical instability.
\citet{LesurKunzFromang2014} confirmed that a third-order Runge-Kutta (RK3) time integrator 
suppresses numerical instabilities in the nonlinear development of Hall-dominated magnetorotational instability 
using \texttt{SNOOPY}, an incompressible pseudo-spectral code.
However, no studies have shown that the RK3 integrator, combined with the simple implementation, provides stable results when using Godunov-type schemes with the constrained transport (CT) methods \citep{EvansHawley1988ApJ...332..659E},
which is a popular combination used in many MHD simulation codes, such as 
\citet[][\texttt{Athena++}]{Stone2020ApJS..249....4S}, 
\citet[][\texttt{RAMSES}]{FromangHennebelleTeyssier2006},
\citet[][\texttt{Pluto}]{Mignone2007ApJS..170..228M},
\citet[][\texttt{Enzo}]{Collins2010ApJS..186..308C}.

Another issue with the simple implementation is that 
it does not provide numerical dissipation because
the Hall electric field is oriented 
perpendicular to the electric current.
This lack of numerical dissipation causes serious problems 
in magnetic reconnection driven by the Hall effect \citep[e.g.,][]{Mandt1994GeoRL..21...73M}.

Several numerical methods have been proposed to suppress numerical 
instabilities caused by the Hall effect.
\citet{Toth2008JCoPh.227.6967T} 
and \citet{LesurKunzFromang2014} suggested modifying 
the signal speeds in the Harten-Lax-van Leer \citep[HLL,][]{HartenLaxvanLeer1983} numerical fluxes
by considering the phase speed of whistler waves to estimate 
the signal speeds of the two characteristics.
This method is called Hall-HLL and 
is widely used in numerical simulations 
of star and planet formation \citep{Bethune2017A&A...600A..75B,BaiStone2017ApJ...836...46B,Marchand2018A&A...619A..37M}.
\citet{Marchand2019AA...631A..66M} proposed a 
modified Hall-HLL method to 
improve the conservation of angular momentum in collapsing dense molecular cloud cores.

Alternative approaches for modifying time integration methods
have also been proposed.
\citet{Falle2003MNRAS.344.1210F} demonstrated that an implicit method stabilizes Hall-MHD.
Furthermore,
\citet{Toth2008JCoPh.227.6967T} showed that an implicit Hall-MHD scheme is stable even without 
modifications to the numerical fluxes.
\citet{OSullivan2006MNRAS.366.1329O,OSullivan2007MNRAS.376.1648O} 
found that the Hall-MHD stability is achieved when 
the magnetic field is updated using a dimensionally split method \citep[also see][]{Bai2014ApJ...791..137B}.

Another approach involves introducing artificial resistivity 
into the induction equation.
For instance, 
hyper-resistivity is added to damp whistler waves 
with wavelengths comparable to the grid scale in 
numerical simulations of magnetic reconnection involving the Hall effect
\citep[e.g.,][]{Birn2001JGR...106.3715B,Ma2001JGR...106.3773M,ChaconKnoll2003JCoPh.188..573C,Vigano2021CoPhC.26508001V}.
However, an appropriate choice of the hyper-resistivity 
coefficient has not been thoroughly examined.
Recently, \citet{Zier2024MNRAS.527.8355Z} proposed 
a method in which 
the diffusion coefficient of Ohmic resistivity is 
artificially increased to stabilize the schemes.

As mentioned earlier, 
various implementations of the Hall effect have been proposed in the literature.  
However, comprehensive comparisons of their stability and accuracy have not yet been conducted.
In this paper, we compare the results of Hall-HLL \citep{LesurKunzFromang2014}, 
a modified version of Hall-HLL \citep{Marchand2019AA...631A..66M}, and 
hyper-resistivity \citep{Birn2001JGR...106.3715B}.

This paper is organized as follows:
Section \ref{sec:basic} reviews the basic properties of Hall-MHD and 
describe the implementations considered in this paper.
Section \ref{sec:experiments} presents numerical experiments.

\section{Basic Properties of Hall-MHD and Its Numerical Implementations} \label{sec:basic}
\subsection{Basic Equations} \label{sec:basic equation}

The basic equations of non-ideal MHD are given by 
\begin{equation}
    \frac{\partial \rho}{\partial t} + \frac{\partial \rho v_i}{\partial x_i} = 0,
    \label{eoc}
\end{equation}
\begin{equation}
    \frac{\partial \rho {v}_i}{\partial t} + 
    \frac{\partial }{\partial x_j}
    (\rho v_i v_j + {\cal T}_{ij})
    = 0,
\end{equation}
\begin{equation}
    \frac{\partial E}{\partial t} + \frac{\partial }{\partial x_i} 
    \left\{ E v_i + {\cal T}_{ij} v_j + (\bm{E}_\mathrm{ni}\times \bm{B})_i \right\} = 0,
\end{equation}
and 
\begin{equation}
    \frac{\partial \bm{B}}{\partial t}
    + \bm{\nabla} \times (-\bm{v}\times \bm{B} + \bm{E}_\mathrm{ni}) = 0,
    \label{induc}
\end{equation}
where $\rho$ is the density, $\bm{v}$ is the velocity, $\bm{B}$ is the magnetic field, 
$P$ is the pressure, 
and $E=\rho \bm{v}^2/2 + P/(\gamma-1) + \bm{B}^2/8\pi$ represents the 
total energy per unit volume.
${\cal T}$ denotes the stress tensor, given by 
\begin{equation}
    {\cal T}_{ij} = \left( P + \frac{B^2}{8\pi}\right)\delta_{ij}
    - \frac{B_iB_j}{4\pi}.
\end{equation}
$\bm{E}_\mathrm{ni}$ represents the electric field resulting 
from non-ideal MHD effects, expressed as
\begin{equation}
    \bm{E}_\mathrm{ni} = \frac{4\pi}{c}\left(\etaO \bm{J}  + 
    \etaH \frac{\bm{J} \times\bm{B}}{|\bm{B}|}
    + \etaA \bm{J}_\perp 
\right),
\end{equation}
$\bm{J}=(c/4\pi)\bm{\nabla}\times \bm{B}$ is the electric current
density, and
$\bm{J}_\perp = \bm{J} - (\bm{J}\cdot\bm{B})/|\bm{B}|$ is the $\bm{J}$ components 
perpendicular to the local magnetic field direction.
$\etaO$, $\etaH$, and $\etaA$ correspond to 
the diffusion coefficients for 
Ohmic resistivity, the Hall effect, and ambipolar diffusion, respectively.
In this study, Ohmic resistivity and ambipolar diffusion are not considered 
because the focus is on implementations that ensure stability and accuracy even with $\etaO=\etaA=0$.


\subsection{Review of Properties of Linear Waves in Hall MHD}\label{sec:linearanalysis}
Although linear Hall-MHD wave propagation tests 
are widely used to evaluate various methods,
most studies focus primarily on incompressible waves 
propagating along the unperturbed magnetic field.
Thus, it remains unclear whether these methods can 
accurately capture other types of linear waves.
In this study, we assess performance of 
various methods on all linear waves in Section \ref{sec:linearwaves}.
This section provides a brief review of the physical properties of the linear waves in Hall-MHD
\citep[e.g.,][]{HameiriIshizawaIshida2005PhPl...12g2109H}.

We consider a uniform, static gas with a density of $\rho_0$ and a uniform magnetic field of $\bm{B}_0$ 
as the unperturbed state.
The sound speed and Alfv\'en speed in the unperturbed state 
are denoted by $c_\mathrm{s}$ and $c_\mathrm{A}$, respectively.
We analyze perturbations proportional 
to $e^{i(\bm{k}\cdot  \bm{x}-\omega t)}$, where $\bm{k}$ is the wavenumber vector 
and $\omega$ is the angular frequency.
Linearizing Equations (\ref{eoc})-(\ref{induc}) with $\etaO=\etaA=0$ yields
the following dispersion relation:
\begin{eqnarray}
    \left(\frac{\omega}{c_\mathrm{A}k}\right)^6 &-& 
    \left\{ \tilde{c}_\mathrm{s}^2 + 1+\cos^2\theta + 
    (kL_\mathrm{H})^2\cos^2\theta\right\}
     \left(\frac{\omega}{c_\mathrm{A}k}\right)^4 \nonumber \\
    &+& \left\{
    \tilde{c}_\mathrm{s}^2
    (k\LH)^2\cos^2\theta 
    + (2\tilde{c}_\mathrm{s}^2 + 1)
  \cos^2\theta
    \right\}
     \left(\frac{\omega}{c_\mathrm{A}k}\right)^2 \nonumber \\
    &-& 
    \tilde{c}_\mathrm{s}^2 \cos^4\theta
    =0,
    \label{disp_hallmhd}
\end{eqnarray}
where $\tilde{c}_\mathrm{s} = c_\mathrm{s}/c_\mathrm{A}$, $k=|\bm{k}|$, 
and $\theta$ is the angle between 
$\bm{k}$ and $\bm{B}_0$.
The Hall scale, $L_\mathrm{H}$, is defined as
\begin{equation}
    L_\mathrm{H} = \frac{\etaH}{c_\mathrm{A}}
    \label{kH}
\end{equation}
\citep{PandeyWardle2008}.
For a fully ionized plasma, $L_\mathrm{H}$ corresponds to the ion skin depth.
When $kL_\mathrm{H}>1$, the dispersion relation deviates from that of ideal MHD 
due to the Hall effect.

\begin{figure}[htpb]
    \centering
    \includegraphics[width=8cm]{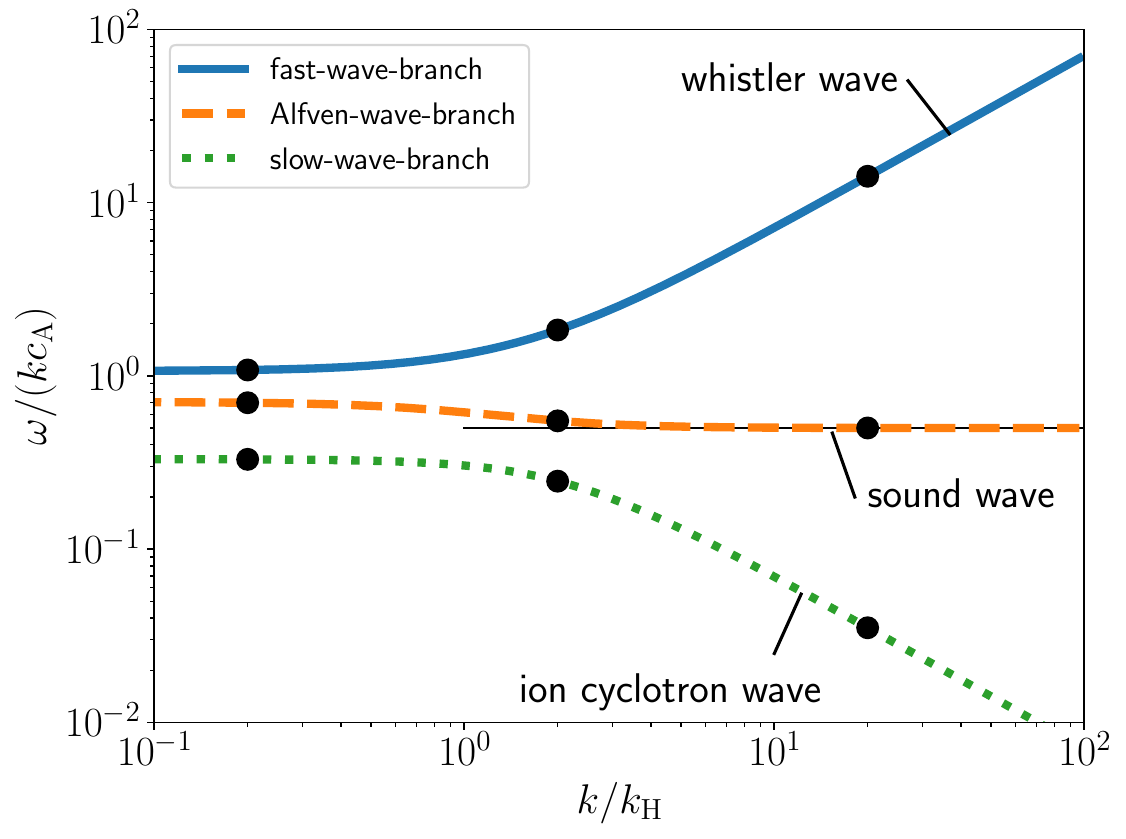}
    \caption{
    Phase speeds of the fast-wave branch, 
    Alfv\'en-wave branch, and slow-wave branch 
    as a function of the wavenumber for $\tcs=1/2$ and $\theta=\pi/4$.
    The horizontal line represents the sound speed.
    The black circles correspond to the six models 
    used in the convergence test conducted in Section \ref{sec:linearwaves}.
    }
    \label{fig:disp_hallmhd}
\end{figure}

Figure \ref{fig:disp_hallmhd} shows the dispersion relation for $\tcs=1/2$ and $\theta=\pi/4$.
As in ideal MHD, three branches appear.
In ascending order of phase speeds, 
they are referred to as the
``slow-wave branch", ``Alfv\'en-wave branch", and ``fast-wave branch".
In the low wavenumber limit, $k\LH\ll 1$,
the slow-wave branch, Alfv\'en-wave branch, and fast-wave 
branch correspond to the slow, Alfv\'en, and fast waves, respectively.

For $kL_\mathrm{H}\gg 1$, the phase speeds of the fast-wave and slow-wave branches 
are no longer constant with respect to the wavenumber 
due to the Hall effect.
The fast-wave branch corresponds to whistler waves,
for which the phase speed is
\begin{equation}
   \frac{\omega}{c_\mathrm{A} k} \sim  k\LH \cos\theta.
\end{equation}
Whistler waves are right-hand circularly polarized.
The magnetic field perturbations oscillate rapidly, while 
the gas remains nearly static.
The velocity perturbations are negligible compared to 
the magnetic field perturbations divided by $\sqrt{4\pi \rho_0}$, 
where $\rho_0$ is the unperturbed density.

For $k\LH\gg 1$, the slow-wave branch asymptotically 
approaches the ion-cyclotron wave, for which the
phase speed is given by 
\begin{equation}
    \frac{\omega}{c_\mathrm{A}k} \sim 
    \left(k\LH\right)^{-1}\cos\theta.
\end{equation}
Ion-cyclotron waves, which are left-hand circularly 
polarized, cannot oscillate at frequencies higher than the 
ion-cyclotron frequency, which is given by $\cA/\LH$ for $\theta=0$.
Unlike whistler waves, 
velocity perturbations dominate relative to
magnetic field perturbations divided by $\sqrt{4\pi \rho_0}$. 

As $k\LH$ increases, 
the Alfv\'en-wave branch transitions from 
an Alfv\'en wave to a sound wave around $k\LH\sim 1$.
Thus, in the Hall-dominated regime ($k\LH\gg 1$),
the compressible mode (sound wave) decouples from
the incompressible modes (whistler and ion-cyclotron waves).

For $\tilde{c}_\mathrm{s}=1/2$,
a mode exchange occurs around $k\LH\sim 1$ between 
the Alfv\'en-wave and slow-wave branches.
The Alfv\'en-wave (slow-wave) 
branch transitions from an incompressible (compressible) state 
to a compressible (incompressble) state.
For gases with $\tilde{c}_\mathrm{s}$ larger than unity, 
a similar mode exchange occurs
between the fast-wave and Alfv\'en-wave branches.

\subsection{Implementations of Hall Effect Considered in this Study}

A simple implementation of the Hall effect 
involves adding 
the Hall electric field to the 
electric field used to update the magnetic field 
without further modifications to the numerical scheme 
(see Section \ref{sec:implementation}).
As mentioned in Section \ref{sec:intro},
a third-order time integrator is conditionally stable against 
infinitesimally small perturbations, whereas
first- and second-order time integrators 
are unstable \citep{Falle2003MNRAS.344.1210F,Kunz2013MNRAS.434.2295K}.
As demonstrated in Section \ref{sec:rk3_hall},
Even with a third-order time integrator, 
nonlinear numerical instabilities arise near
the Nyquist wavelength when perturbation amplitudes are large.

To suppress numerical instabilities,
various implementations have been proposed in the literature.
In this section, in addition to the simple implementation,
we introduce three methods for comparison in this study.
The implementation details of these methods 
in \texttt{Athena++} 
are described in Section \ref{sec:implementation}.

\subsubsection{Hall-HLL}

Since the Hall effect modifies the phase speeds of linear waves (Section \ref{sec:linearwaves}), 
it is natural to modify the Riemann solver used to compute 
the numerical fluxes.
The Hall effect makes the phase speeds of waves to 
be wavenumber-dependent 
(Figure \ref{fig:disp_hallmhd}). 
It is challenging to construct a Riemann solver incorporating
the Hall effect 
because self-similarity no longer holds in the Riemann problem. 
Among various approximate Riemann solvers, the HLL method \citep{HartenLaxvanLeer1983} 
can be applied to any problem by setting signal speeds appropriately.  
\citet{LesurKunzFromang2014} modified the HLL numerical flux 
by incorporating the phase speed of whistler waves. 
This method is referred to as \textsc{Hall-HLL}. 

In the HLL scheme \citep{HartenLaxvanLeer1983}, by 
considering two waves propagating at 
the signal speeds $S_\mathrm{L}$ and $S_\mathrm{R}$ 
from the initial discontinuity at the cell boundary, 
the numerical flux at the cell boundary
is constructed as follows:
\begin{equation}
    \bm{F}_\mathrm{HLL} = \left\{
    \begin{array}{ll}
       \bm{F}_\mathrm{L}  &  S_\mathrm{L}>0\\
       \bm{F}_\mathrm{R}  &  S_\mathrm{R}<0\\
       \displaystyle 
       \frac{ S_\mathrm{R} S_\mathrm{L} (\bm{U}_\mathrm{R} - \bm{U}_\mathrm{L}) 
       + S_\mathrm{R} \bm{F}_\mathrm{L} - S_\mathrm{L}\bm{F}_\mathrm{R}}{ S_\mathrm{R} - S_\mathrm{L}} & \mathrm{otherwise}\\
    \end{array}
    \right.,
    \label{hallhll}
\end{equation}
where $\bm{U}_\mathrm{L}$ and $\bm{U}_\mathrm{R}$ are the conserved variables 
of the left and right states, and 
$\bm{F}_\mathrm{L}$ and $\bm{F}_\mathrm{R}$ are the corresponding fluxes.
Expressions of the signal speeds $S_\mathrm{L}$ and $S_\mathrm{R}$ are replaced with
\begin{equation}
   S_\mathrm{L} = \min(v_\mathrm{L} - c_\mathrm{max,L}, v_\mathrm{R} - c_\mathrm{max,R}) 
   \label{SL}
\end{equation}
and 
\begin{equation}
   S_\mathrm{R} = \max(v_\mathrm{L} + c_\mathrm{max,L}, v_\mathrm{R} + c_\mathrm{max,R}),
   \label{SR}
\end{equation}
respectively \citep{Davis1988}, where $v_\mathrm{L}$ and $v_\mathrm{R}$ 
are the normal velocities of left and right states, respectively.
$c_\mathrm{max,L}$ ($c_\mathrm{max,R}$) is the maximum phase speed 
$c_\mathrm{max}$ of the left (right) state.
In ideal MHD, the phase speed of fast waves $c_\mathrm{f}$ is assigned to $c_\mathrm{max}$.
Note that there are other options for the expressions of 
$S_\mathrm{L,R}$ \citep[e.g.,][]{Einfeldt1988SJNA...25..294E}.

In \textsc{Hall-HLL}, $c_\mathrm{max}$ is given by 
\begin{equation}
    c_\mathrm{max} =  \max(c_\mathrm{f},c_\mathrm{w}(k_\mathrm{max})),
    \label{cmax}
\end{equation}
where $c_\mathrm{w}$ is the phase speed of whistler waves propagating along the magnetic field,
\begin{equation}
    c_\mathrm{w}(k) = \frac{\etaH k}{2} + \sqrt{ \left(\frac{\etaH k }{2}\right)^2 + c_\mathrm{A}^2},
    \label{chall}
\end{equation}
and $k_\mathrm{max}$ is the maximum wavenumber that can be resolved in numerical simulations.

Various values of $k_\mathrm{max}$ were used in 
previous studies.
\citet{LesurKunzFromang2014} and \citet{BaiStone2017ApJ...836...46B} adopted
$k_\mathrm{max}=\Delta x^{-1}$, while 
\citet{Marchand2018A&A...619A..37M} employed
a larger value of $k_\mathrm{max} = \pi \Delta x^{-1}$ that corresponds to 
the Nyquist wavenumber.
In this study, $k_\mathrm{max}=\Delta x^{-1}$ is adopted because
the results of \citet{LesurKunzFromang2014} and \citet{BaiStone2017ApJ...836...46B} demonstrate
that $k_\mathrm{max}=\Delta x^{-1}$ is sufficiently large to suppress numerical instabilities caused by the Hall effect.

The numerical dissipation introduced by \textsc{Hall-HLL} is estimated from the induction equation.
For simplicity, we consider a one-dimensional problem where
the Hall effect dominates ($c_\mathrm{w}(\Delta x^{-1})\gg c_\mathrm{f}$) and 
the flow speed is significantly lower than $c_\mathrm{w}(\Delta x^{-1})$. 
From Equations (\ref{SL}), (\ref{SR}), and (\ref{cmax}),
the signal speeds are given by $S_\mathrm{L} = -c_\mathrm{w}(\Delta x^{-1})$ and 
$S_\mathrm{R} = c_\mathrm{w}(\Delta x^{-1})$.
\citet{Toth2008JCoPh.227.6967T} derived the following equation:
\begin{eqnarray}
\left( \frac{\partial \bm{B}_\perp}{\partial t}\right)_{\text{\sc Hall-HLL}}
&\sim & 
- \frac{c_\mathrm{w}(\Delta x^{-1})\Delta x^3}{8}
\frac{\partial^4 \bm{B}_\perp}{\partial x^4}\nonumber \\
&\sim & 
- \frac{\etaH\Delta x^2}{8}
\frac{\partial^4 \bm{B}_\perp}{\partial x^4},
\label{diss_hallhll}
\end{eqnarray}
where Equation (\ref{chall}) is applied in the limit $\etaH \Delta x^{-1}\gg \cA$.

\subsubsection{Modified Hall-HLL}

\citet{Marchand2018A&A...619A..37M} found that 
the use of {\sc Hall-HLL} significantly violates 
angular momentum conservation in 
simulations of gravitational collapse of dense cores in Cartesian coordinates.

A modified version of {\sc Hall-HLL} was 
proposed by \citet{Marchand2019AA...631A..66M}.
This method is referred to as {\sc Hall-HLLmod} in this paper.
In {\sc Hall-HLLmod}, 
Equation (\ref{cmax}) is exclusively used to 
compute the numerical fluxes for the magnetic field.
For other components of the numerical fluxes 
used to update the hydrodynamic variables 
($\rho,\rho\bm{v},E$),   
the original HLL numerical flux is applied.
It has been found that {\sc Hall-HLLmod} significantly improves
angular momentum conservation in collapsing dense cores.


\subsubsection{Hyper Resistivity}

Hyper resistivity is introduced into the electric field as follows:
\begin{equation}
    \bm{E}_\mathrm{hyp} = - \frac{4\pi}{c}\eta_\mathrm{hyp} \bm{\nabla}^2\bm{J},
    \label{hypresis}
\end{equation}
where $\eta_\mathrm{hyp}$ is a coefficient
\citep{Birn2001JGR...106.3715B, Ma2001JGR...106.3773M}.
The coefficient $\eta_\mathrm{hyp}$ must be sufficiently 
large to suppress 
numerical instabilities but small enough to ensure accurate results.

To ensure that the time step limitation imposed by hyper-resistivity is less restrictive than that imposed by the Hall effect for any $\Delta x$ and $\etaH$, 
the coefficient is defined as follows:
\begin{equation}
    \eta_\mathrm{hyp} = C_{\mathrm{hyp}} \etaH \Delta x^2,
    \label{etahyp}
\end{equation}
where $C_{\mathrm{hyp}}$ is a free parameter.
\citet{ChaconKnoll2003JCoPh.188..573C} employed
a similar coefficient to develop 
a two-dimensional implicit Hall-MHD solver.
A possible range of $C_\mathrm{hyp}$ is estimated in Section \ref{sec:Chypmin}.

In a one-dimensional problem, 
the dissipation term due to the hyper resistivity is expressed as 
\begin{eqnarray}
\left( \frac{\partial \bm{B}_\perp}{\partial t}\right)_{\text{\sc hyp-resis}}
= - C_\mathrm{hyp}\etaH\Delta x^2
\frac{\partial^4 \bm{B}_\perp}{\partial x^4} 
\label{diss_hyp}
\end{eqnarray}
in the long wavelength limit.
Comparison between Equations (\ref{diss_hallhll}) and (\ref{diss_hyp}) shows that 
their dissipation terms are identical when
$C_\mathrm{hyp}=0.125$.

\subsection{Implementations of Hall Effect in Athena++}\label{sec:implementation}

We implement the four methods listed in Table \ref{tab:methods} 
in \texttt{Athena++} \citep{Stone2020ApJS..249....4S}.
For all methods, we use the third-order strong stability 
preserving Runge-Kutta time integrator that is referred to 
as RK3 \citep[][Equation (3.1)]{Gottlieb2009} and 
the piecewise linear spatial reconstruction with the 
van-Leer limiter \citep{vanLeer1974}.
Except for {\sc Hall-HLL} and {\sc Hall-HLLmod}, 
the HLLD numerical flux without modifying the signal speeds is used \citep{MiyoshiKusano2005JCoPh.208..315M}.

\begin{table*}[htpb]
\caption{List of the implementations evaluated in this paper.}
\begin{center}
\begin{tabular}{|c|c|c|}
 \hline
 method name & description of implementation \\
 \hline
 \hline
 {\sc HLLD}  & 
             \begin{minipage}{0.8\textwidth} 
                   \vspace{1mm} 
             No modifications are made in \texttt{Athena++} 
             except for the addition of the Hall term 
             to the electric field used to update the magnetic field in the constrained transport method.  
             \end{minipage}
                    \\
 \hline
 {\sc Hall-HLL} & \begin{minipage}{0.8\textwidth} 
                   \vspace{1mm} 
                   The Hall-HLL numerical flux is used instead of {\sc HLLD}
                   (Equation (\ref{hallhll}))  \citep{LesurKunzFromang2014}
                   \end{minipage} \\
 \hline
 {\sc Hall-HLLmod} & \begin{minipage}{0.8\textwidth} 
                   \vspace{1mm} 
                   The Hall-HLL numerical flux 
                   is used to update the magnetic field.
                   (Equation (\ref{hallhll})).
                   For other variables, the numerical 
                   flux are calculated by replacing $c_\mathrm{max}$ by $c_\mathrm{f}$ \citep{Marchand2019AA...631A..66M}.
                   \end{minipage} \\
 \hline 
 {\sc Hyp-Resis} & \begin{minipage}{0.8\textwidth} 
                   \vspace{1mm} 
                   A fourth-order hyper resistivity is added into the electric field 
                   (Equation (\ref{hypresis})) \citep{Birn2001JGR...106.3715B}.
                   \end{minipage} \\
                   \hline
\end{tabular}
\end{center}
\label{tab:methods}
\end{table*}

\texttt{Athena++} employs a staggered grid for the CT scheme. 
The conserved hydrodynamical variables 
$(\rho,\rho\bm{v},E)$ are averaged within 
the cell volume and defined 
at the cell volume center. 
The normal components of the magnetic field $\bm{B}$ 
are averaged on cell surfaces and are defined at the cell surface center.
In this paper, we consider only Cartesian coordinates.

The cell center coordinates are denoted by
$(x_i,y_j,z_k)$, where $i,j,k$ represent the discrete cell indices. 
A cell-centered variable, $U$, at $(x_i,y_j,z_k)$ is expressed as $U_{i,j,k}$.
The positions of the cell surface between $(i,j,k)$-th and 
$(i+1,j,k)$-th cells are denoted as $(x_{i+1/2},y_j,z_k)$.
The $x$-component of the magnetic field defined at $(x_{i+1/2},y_j,z_k)$ 
is $(B_x)_{i+1/2,j,k}$.
Similarly, 
the $y$- and $z$-components of the magnetic field are denoted as 
$(B_y)_{i,j+1/2,k}$, and $(B_z)_{i,j,k+1/2}$, respectively.

The hydrodynamical variables $(\rho,\rho\bm{v},E)$ are 
updated by 
computing the numerical fluxes using a Riemann solver.
In deriving the numerical fluxes, 
the cell-centered magnetic fields are computed as follows:
\begin{eqnarray}
 & \displaystyle 
 (B_x)_{i,j,k} = \frac{1}{2}\left\{(B_x)_{i-\half,j,k} + (B_x)_{i+\half,j,k}\right\} \nonumber \\
 & \displaystyle 
 (B_y)_{i,j,k} = \frac{1}{2}\left\{(B_y)_{i,j-\half,k} + (B_y)_{i,j+\half,k}\right\}\\
 & \displaystyle 
 (B_z)_{i,j,k} = \frac{1}{2}\left\{(B_z)_{i,j,k-\half} + (B_z)_{i,j,k+\half}\right\}\nonumber.
 \label{face_to_center}
\end{eqnarray}

To update the surface-centered components of the magnetic field, 
the electric field components at the cell edges are 
computed, namely, $(E_x)_{i,j+1/2,k+1/2}$, $(E_y)_{i+1/2,j,k+1/2}$, $(E_z)_{i+1/2,j+1/2,k}$.
The electric fields from ideal MHD ($-\bm{v}\times\bm{B}$)
are calculated using the method 
proposed by \citet{GardinerStone2005JCoPh.205..509G,GardinerStone2008JCoPh.227.4123G}.

\subsubsection{Inclusion of the Hall Term in the Induction Equation}
Similar to Ohmic diffusion and ambipolar diffusion, 
the Hall effect is incorporated into the cell-edge 
electric fields.
Discretized expressions of
the electric currents, defined at the cell edges, are given by 
\begin{eqnarray}
   (J_x)_{i,j+\half,k+\half} &=& 
   \frac{(B_z)_{i,j+1,k+\half} - (B_z)_{i,j,k+\half}}{\Delta y} \nonumber \\  
   &-& 
   \frac{(B_y)_{i,j+\half,k+1} - (B_y)_{i,j+\half,k}}{\Delta z}.   
\end{eqnarray}
The remaining components,
$(J_y)_{i+1/2,j,k+1/2}$ and 
$(J_z)_{i+1/2,j+1/2,k}$, are computed in the same manner.

The $x$-component of the electric field induced by the Hall effect is given by 
\begin{eqnarray}
    (E_x)_{i,j+\half,k+\half} &=& 
    \frac{(\etaH)_{i,j+\half,k+\half}}{|\bm{B}_{i,j+\half,k+\half}|}
    \nonumber \\
    & \times &\left[
       (J_y)_{i,j+\half,k+\half} 
       (B_z)_{i,j+\half,k+\half} 
       \right. \nonumber \\
       && 
       \left.
       - 
       (J_z)_{i,j+\half,k+\half} 
       (B_y)_{i,j+\half,k+\half} 
    \right],
    \label{Exhall}
\end{eqnarray}
The quantities in Equation (\ref{Exhall}) 
defined at the cell edge $(x_i,y_{j+1/2},z_{k+1/2})$ are 
calculated as follows:
\[
(\etaH)_{i,j+\half,k+\half} = \frac{1}{4}
\sum_{kl=0}^1 \sum_{jl=0}^1 (\etaH)_{i,j+jl,k+kl},
\]
\[
(B_x)_{i,j+\half,k+\half} = \frac{1}{8}
\sum_{il=0}^1\sum_{jl=0}^1\sum_{kl=0}^1
(B_x)_{i-\half+il,j+il,k+kl},
\]
\[
(B_y)_{i,j+\half,k+\half} = \frac{1}{2}\left\{
(B_y)_{i,j-\half,k}
+ (B_y)_{i,j+\half,k}
\right\},
\]
\[
(B_z)_{i,j+\half,k+\half} = \frac{1}{2}\left\{
(B_z)_{i,j,k-\half}
+ (B_z)_{i,j,k+\half}
\right\},
\]
\[
(J_y)_{i,j+\half,k+\half} = \frac{1}{4}
\sum_{il=0}^1\sum_{jl=0}^1 (J_y)_{i-\half+il,j+jl,k+\half},
\]
and
\[
(J_z)_{i,j+\half,k+\half} = \frac{1}{4}
\sum_{il=0}^1\sum_{kl=0}^1 (J_z)_{i-\half+il,j+\half,k+kl}.
\]

\subsubsection{Implementations of Hyper-resistivity}

The hyper-resistivity coefficient $\eta_\mathrm{hyp}$ shown 
in Equation (\ref{etahyp}) can be applied only when $\Delta x=\Delta y=\Delta z$.
To satisfy the stability condition, $\Delta x^2$ in   Equation (\ref{etahyp}) is replaced with $\min(\Delta x^2,\Delta y^2,\Delta z^2)$.
The discretized form of hyper resistivity is given as follows: 
\begin{eqnarray}
    (E_{x,\mathrm{hyp}})_{i,j+\half,k+\half} 
   & = &
   C_\mathrm{hyp}(\etaH)_{i,j+\half,k+\half} \nonumber \\
   & & 
   \min(\Delta x^2,\Delta y^2,\Delta z^2)
   \nonumber \\
  &&  \times  \Delta_\mathrm{L} (J_x)_{i,j+\half,k+\half} ,
\end{eqnarray}
where $\Delta_\mathrm{L}$ represents the 
discrete Laplacian operator, defined as: 
\begin{eqnarray}
    \Delta_\mathrm{L} Q_{i,j,k}
  & = & \frac{1}{\Delta x^2} \left(
    Q_{i-1,j,k}
  -2 Q_{i,j,k}
  + Q_{i+1,j,k}
    \right)\nonumber \\
  & + & \frac{1}{\Delta y^2} \left(
    Q_{i,j-1,k}
  -2 Q_{i,j,k}
  + Q_{i,j+1,k}
    \right)\nonumber \\
  & + & \frac{1}{\Delta z^2} \left(
    Q_{i,j,k-1}
  -2 Q_{i,j,k}
  + Q_{i,j,k+1}
    \right)
\end{eqnarray}
assuming a uniform grid spacing.

\subsubsection{Time Step Constraint due to the Hall Effect}\label{sec:CFL}

Based on the von Neumann stability analysis 
(Appendix \ref{app:vonNeumann}), 
the time step constraint imposed by the Hall effect is expressed as 
\begin{equation}
    \Delta t_\mathrm{H} = 
    \frac{C_\mathrm{H}}{4}\sqrt{\frac{3}{d}}
    \min_{i,j,k}\left\{
    \frac{\min(\Delta x^2, \Delta y^2, \Delta z^2)}{(\etaH)_{i,j,k}}
    \right\},
\end{equation}
where $d$ represents the spatial dimension.
In this study, $C_\mathrm{H}$ is set to 0.8.

\section{Numerical Experiments}\label{sec:experiments}

\subsection{Stability of Hall-MHD with {HLLD}}\label{sec:rk3_hall}

\citet{Kunz2013MNRAS.434.2295K} demonstrated that 
Hall-MHD with third-order time integrators is 
conditionally stable in the absence of dissipation,
whereas second-order time integrators lead to 
numerical instability for any $\Delta t$ 
\citep[see also][]{Falle2003MNRAS.344.1210F}.
In this section, we analyze the stability of the RK3 integrator 
through whistler-wave propagation tests. 

In the whistler-wave propagation test, 
most studies focus on waves propagating exclusively
along the unperturbed magnetic field.
To explore a more general situation, 
we examine a whistler wave propagating at an angle relative to the wavenumber vector $\bm{k}$.

The wavenumber vector is inclined relative to the grid 
\citep{GardinerStone2008JCoPh.227.4123G, Mignone2010JCoPh.229.5896M}. 
The wavenumber vector $\bm{k}$ is set to $2\pi(1,2,2)/3$.
The unit vector along the $x$-axis is converted into 
$\bm{k}/|\bm{k}|$ using the rotation matrix:
\begin{equation}
{\cal R} = 
\left(
    \begin{array}{ccc}
       \cos\theta_1 & -\sin\theta_1 & 0 \\
       \sin\theta_1 & \cos\theta_1 & 0 \\ 
       0 & 0 &  1
    \end{array}
\right)
\cdot
\left(
    \begin{array}{ccc}
       \cos\theta_2 & 0 & -\sin\theta_2  \\
       0 & 1 & \\
       \sin\theta_2 & 0 & \cos\theta_2  \\ 
    \end{array}
\right),
\end{equation}
where $\tan\theta_1 = 2$ and $\tan\theta_2 = 2/\sqrt{5}$.

The coordinates $\bm{\xi}=(\xi,\eta,\zeta)$ are defined as ${\cal R}\bm{x}$,
aligning the wavenumber vector $\bm{k}$ to the $\xi$-axis.
The unperturbed magnetic field $\bm{B}_0$ is tilted 
by $\theta_\mathrm{B}$ with respect to the $\xi$-axis in the ($\xi,\eta$) plane, where
$\bm{B}_0 = (B_{\xi0}\cos\theta_\mathrm{B},B_{\eta0}\sin\theta_\mathrm{B},0)$.

The Hall coefficient $\etaH$ is set so that $kL_\mathrm{H}=2\times 10^4\pi$.
For $k\LH\gg 1$, the gas remains static in the fast-wave branch,
which corresponds to whistler waves.
Thus, perturbations are introduced exclusively in the magnetic field as follows:
\begin{eqnarray}
   \delta B_\xi=0,\;\; \delta B_\eta = A\sin(k\xi),\;\;\delta B_\zeta = A\cos(k\xi),
\end{eqnarray}
where $A$ represents the initial amplitude and $k=|\bm{k}|$.
The dispersion relation of the whistler waves is 
\begin{equation}
    \frac{\omega}{k} = \frac{\etaH k\cos\theta_\mathrm{B}}{2}
    + \sqrt{ \left(\frac{\etaH k\cos\theta_\mathrm{B}}{2}\right)^2 
    + c_\mathrm{A}^2 \cos^2\theta_\mathrm{B}}.
    \label{disp_whistler_oblique}
\end{equation}

The computational box spans $0\le x \le 2$ and $0\le y,z\le 1$, 
and is discretized into $N\times (N/2)\times (N/2)$ cells.
Periodic boundary conditions are imposed in all directions.
Numerical instabilities caused by the Hall effect are expected to develop around the Nyquist wavelength.
A small value of $N=16$ is used to promote numerical instabilities.

We consider two values for the initial amplitude ($A=10^{-3}$ and $10^{-1}$) and 
analyze how the propagation of the whistler waves depends on $A$ and $\theta_\mathrm{B}$.

\begin{figure}[htpb]
    \centering
    \includegraphics[width=8cm]{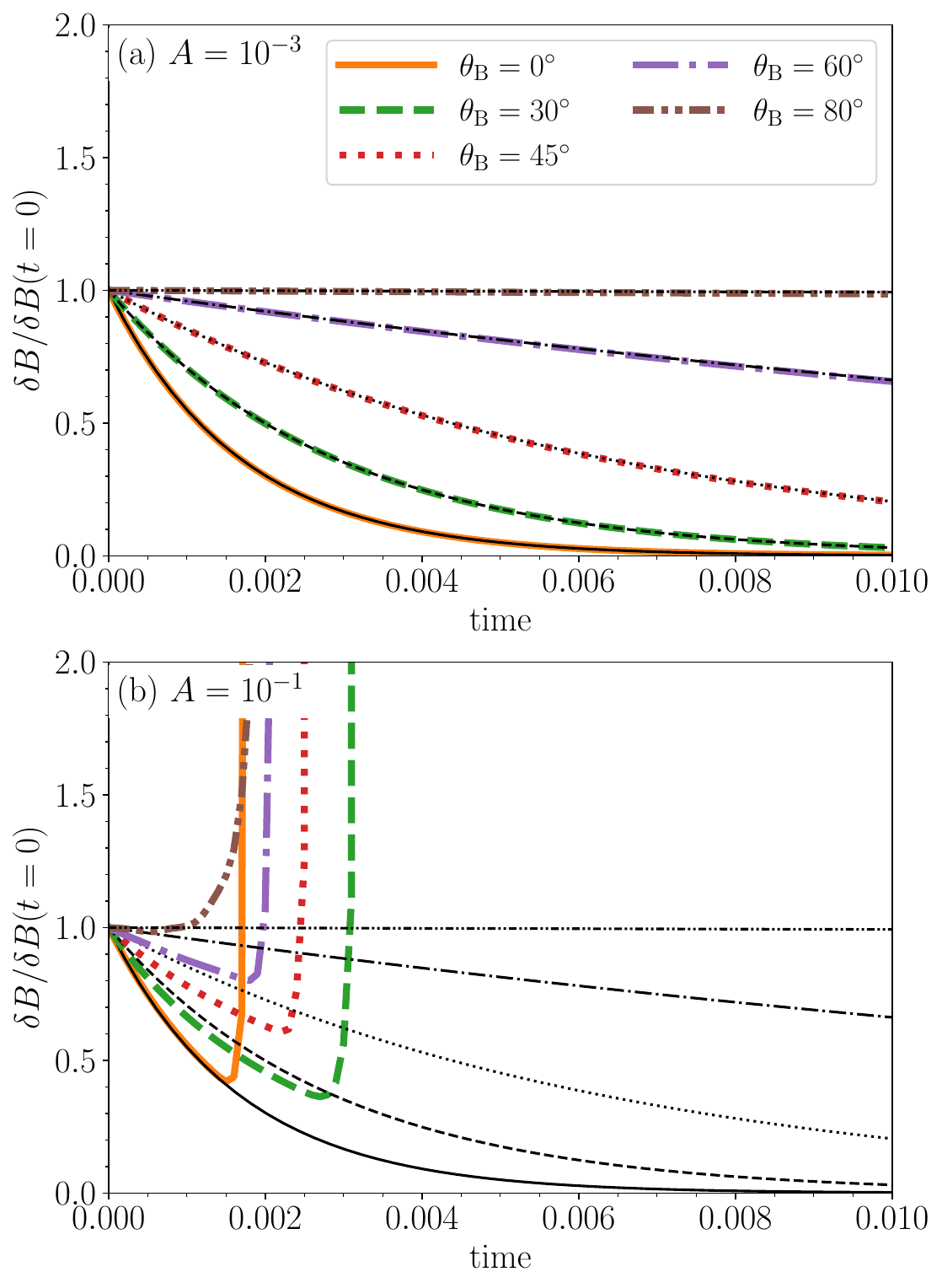}
    \caption{
    Time evolution of 
    $\delta B = (\langle \delta B_x^2\rangle+
    \langle\delta B_y^2\rangle+\langle\delta B_z^2\rangle)^{1/2}$
    for (a) $A=10^{-3}$ and (b) $A=10^{-1}$.
    The results with 
    five different $\theta_\mathrm{B}$ ($0^\circ$, $30^\circ$, $45^\circ$, $60^\circ$, and $80^\circ$) are shown.
    The thin lines represent the predictions from the 
    von Neumann stability analysis presented in 
    Appendix \ref{app:vonNeumann}.
    Differences in linestyle indicate variations 
    in $\theta_\mathrm{B}$.
    }
    \label{fig:disp-whis}
\end{figure}

Figure \ref{fig:disp-whis}a presents the results for the smaller amplitude, $A=10^{-3}$.
For all $\theta_\mathrm{B}$, 
the time evolution of $\delta B = (\langle \delta B_x^2\rangle+
\langle\delta B_y^2\rangle+\langle\delta B_z^2\rangle)^{1/2}$
is consistent with the predictions from the von Neumann 
stability analysis presented in Appendix \ref{app:vonNeumann}.
Minor discrepancies between the numerical results and theoretical predictions are
attributed to numerical dissipation introduced by the HLLD solver that 
is not considered in the von Neumann stability analysis.

\begin{figure}[htpb]
    \centering
    \includegraphics[width=8cm]{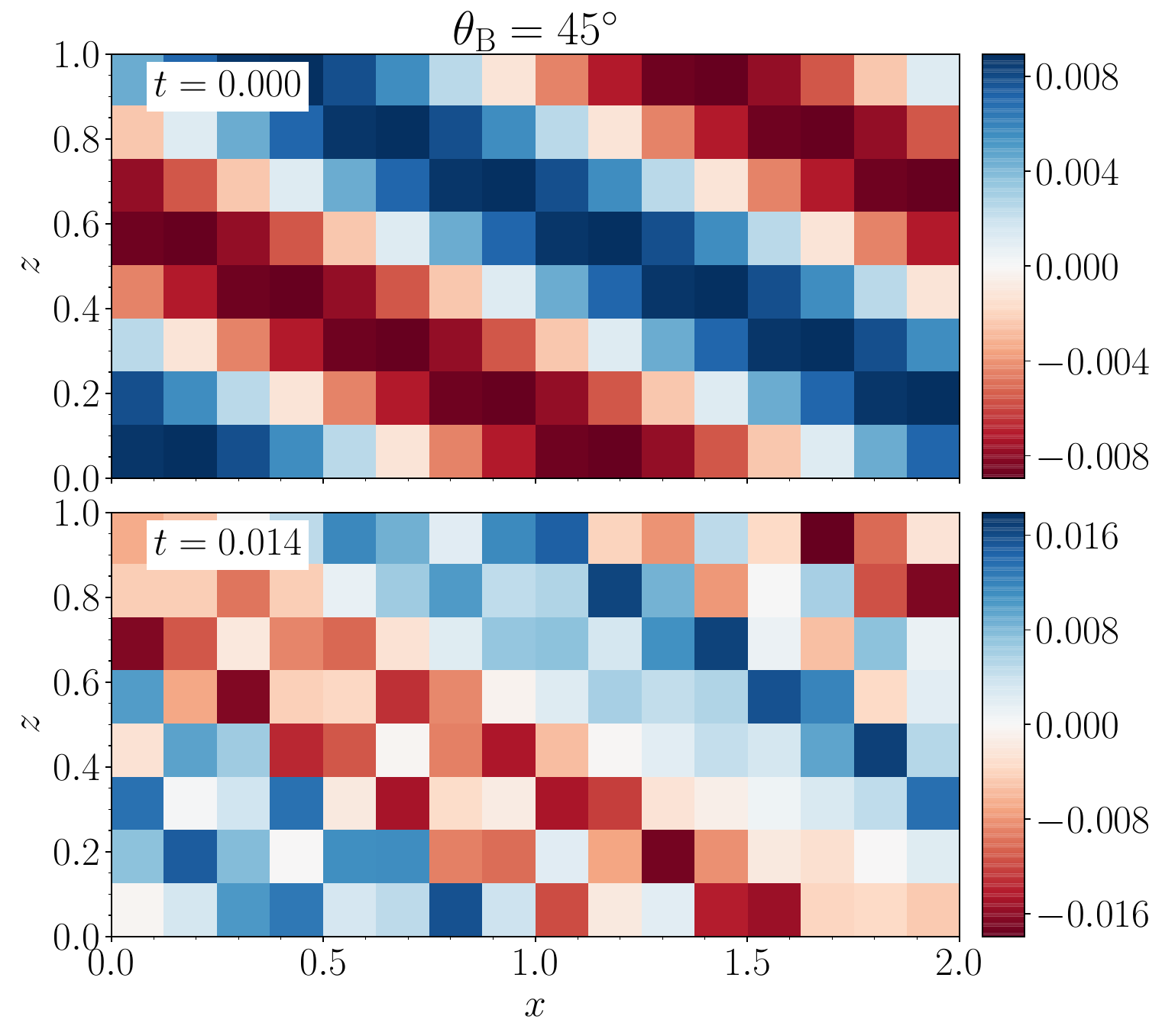}
    \caption{Two-dimensional slice of $\delta B_z$ at $y=0.5$.
    The top panel shows the initial condition, and 
    the bottom panel shows the $\delta B_z$ map when
    the amplitude of $\delta B_z$ has increased by a factor of two.
    }
    \label{fig:whistler_slice}
\end{figure}

When $A$ increases from $10^{-3}$ to $10^{-1}$, its behavior changes significantly.
Numerical instabilities occur suddenly and 
$\delta B$ increases rapidly at the grid scale for all the cases.
Figure \ref{fig:whistler_slice} compares the $B_z$ maps at $t=0$ and shortly after the onset of 
the numerical instability.
Grid-scale fluctuations in the magnetic field increase.

\subsection{Turbulent Magnetic Field}\label{sec:whitenoise}

In this test problem, we first measure the growth rate of the numerical instability induced by the Hall effect in 
Section \ref{sec:growth}.
Using both the growth rate and the results of the von Neumann stability analysis of Hall-MHD with hyper resistivity (Appendix \ref{app:vonNeumann}), we determine an appropriate range of $C_\mathrm{hyp}$ in Section \ref{sec:Chypmin}.
The implementations listed in Table \ref{tab:methods} are compared in Section \ref{sec:turb_comp}

Numerical instabilities caused by the Hall effect 
typically occur around the Nyquist wavelength.
To evaluate the performance of the methods listed in 
Table \ref{tab:methods}, 
we initialize a uniform, static gas ($\rho=1$, $P=1/\gamma$, $\bm{v}=0$) 
with turbulent magnetic fields that exhibit white noise, where $\gamma=5/3$.
No net magnetic field is present.

To ensure that $\bm{\nabla}\cdot \bm{B}=0$ within round-off errors, 
the initial magnetic field is derived from
a turbulent vector potential field, with a power spectrum designed 
to generate white noise in the face-centered magnetic field fluctuation.
The initial amplitude of the magnetic field perturbation is 
defined as $\sqrt{\langle \delta \bm{B}^2\rangle}=\sqrt{4\pi}$, where 
$\langle Q \rangle$ denotes the volume average of $Q$.
Note that the results do not depend on the field strength because 
the induction equation, which considers only the Hall electric field, is 
linear with respect to $\bm{B}$ when $\etaH$ is constant
as long as the phase speed of the whistler wave at the grid scale is significantly larger than 
the Alfv\'en speed and sound speed.
 
In addition, as shown in Section \ref{sec:linearanalysis},
in whistler waves, magnetic field perturbations dominate over 
other perturbations.
In other words, the gas is almost static during the development of the magnetic field.

The computational domain spans $0\le x,y,z\le L$ and is discretized into $32^3$ cells, with 
periodic boundary conditions applied in all directions. 

\subsubsection{Measuring a Growth Rate of the Numerical Instabilities due to Hall Effect}\label{sec:growth}

Before presenting the results of the four implementations, 
we estimate the growth rate of the numerical instabilities caused 
by the Hall effect.
In this test, {\sc HLLD} is used.

Since the numerical instabilities primarily develop around the Nyquist wavelength, 
the characteristic timescale is expected to correspond to 
the crossing time of the whistler wave
across this wavelength.
\begin{equation}
t_\mathrm{w} = \frac{2\Delta x^2}{\pi \eta_\mathrm{H}},
\label{tw}
\end{equation}
where we use the phase speed of the whistler wave at the Nyquist wavelength $\etaH \pi/\Delta x$.
Thus, the growth rate $\sigma$ can be parametrized as
\begin{equation}
    \sigma_\mathrm{inst} = C_\sigma\frac{\eta_\mathrm{H}}{\Delta x^2},
    \label{sigma}
\end{equation}
where $C_\sigma$ is a parameter that is determined by the numerical experiment shown below.

Figure \ref{fig:sig} shows the time evolution of $\sqrt{\langle \delta B^2\rangle}$ for 
the two parameter sets, $(\eta_\mathrm{H}=10^4,L=4)$ and $(\eta_\mathrm{H}=10^3,L=8)$.
The two lines are almost identical when the normalized time $\etaH t/\Delta x^2$ is taken as the horizontal axis.
This clearly shows that the growth rate $\sigma_\mathrm{inst}$ is proportional to $t_\mathrm{w}^{-1}\sim \etaH \Delta x^{-2}$.
Fitting the results with $\exp(\sigma_\mathrm{inst} t)$ yields $C_\sigma = 0.5$.

\begin{figure}[htpb]
    \centering
    \includegraphics[width=8.5cm]{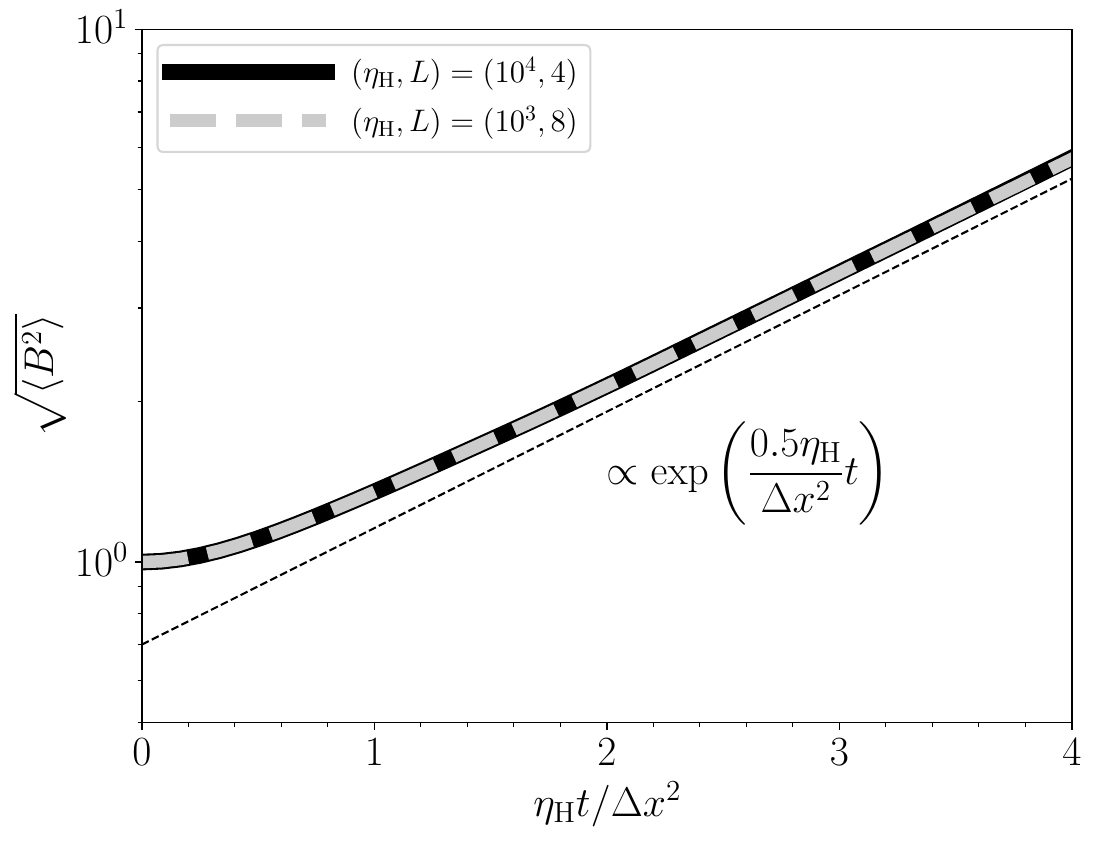}
    \caption{
Time evolution of $\sqrt{\langle \delta B^2\rangle}$ for 
the two parameter sets $(\eta_\mathrm{H}=10^4,L=4)$ and $(\eta_\mathrm{H}=10^3,L=8)$.
  The thin dashed line represents the estimated growth rate.
    }
    \label{fig:sig}
\end{figure}

\subsubsection{Appropriate Range of the Hyper Resistivity Coefficient }\label{sec:Chypmin}

The minimum value of $C_\mathrm{hyp}$ is determined by the condition 
that the damping rate caused by the hyper resistivity  exceeds 
the growth rate $\sigma$ of the numerical instability. 

By linearizing the discretized induction equation solved in \texttt{Athena++},
we obtain a damping rate of $C_\mathrm{hyp}\etaH 16d_\mathrm{Ny}^2\Delta x^{-2}$, considering 
only the hyper-resistivity in the linear analysis, 
assuming cubic cells 
$(\Delta x = \Delta y=\Delta z)$, where
$d_\mathrm{Ny}$ denotes the number of directions containing 
Nyquist wavelength fluctuations.
Therefore, $C_\mathrm{hyp}$ must meet the condition
\begin{equation}
    C_\mathrm{hyp} > C_\mathrm{hyp,min} \sim 0.03 d_\mathrm{Ny}^{-2}.
    \label{Chypmin}
\end{equation}
Note that since $d_\mathrm{Ny}=1$ is adopted to obtain a stricter condition for $C_\mathrm{hyp}$, 
$C_\mathrm{hyp,min}\sim 0.03$ should be regarded as an upper limit of $C_\mathrm{hyp,min}$.

The maximum value of $C_\mathrm{hyp}$ is determined by 
the condition that
the time step constraint caused by hyper resistivity, 
$\Delta t_\mathrm{hyp}$, is 
greater than $\Delta t_\mathrm{H}$.
From the von Neumann stability analysis, 
$\Delta t_\mathrm{hyp} \propto d^2\Delta x^2/\etaH$ can be 
obtained.
By conducting numerical experiments on the turbulent magnetic field with 
different $C_\mathrm{H}$, $d$, and $C_\mathrm{hyp}$, we found 
that $C_\mathrm{hyp}$
must satisfy the following condition:
\begin{equation}
    C_\mathrm{hyp} < C_\mathrm{hyp,max} = 
    0.09 \left(\frac{d}{3}\right)^{-3/2} \left(\frac{C_\mathrm{H}}{0.8}\right)^{-1} 
\end{equation}
in order for $\Delta t_\mathrm{hyp}$ not to affect the CFL condition (Section \ref{sec:CFL}).


\begin{figure}[htpb]
    \centering
    \includegraphics[width=8cm]{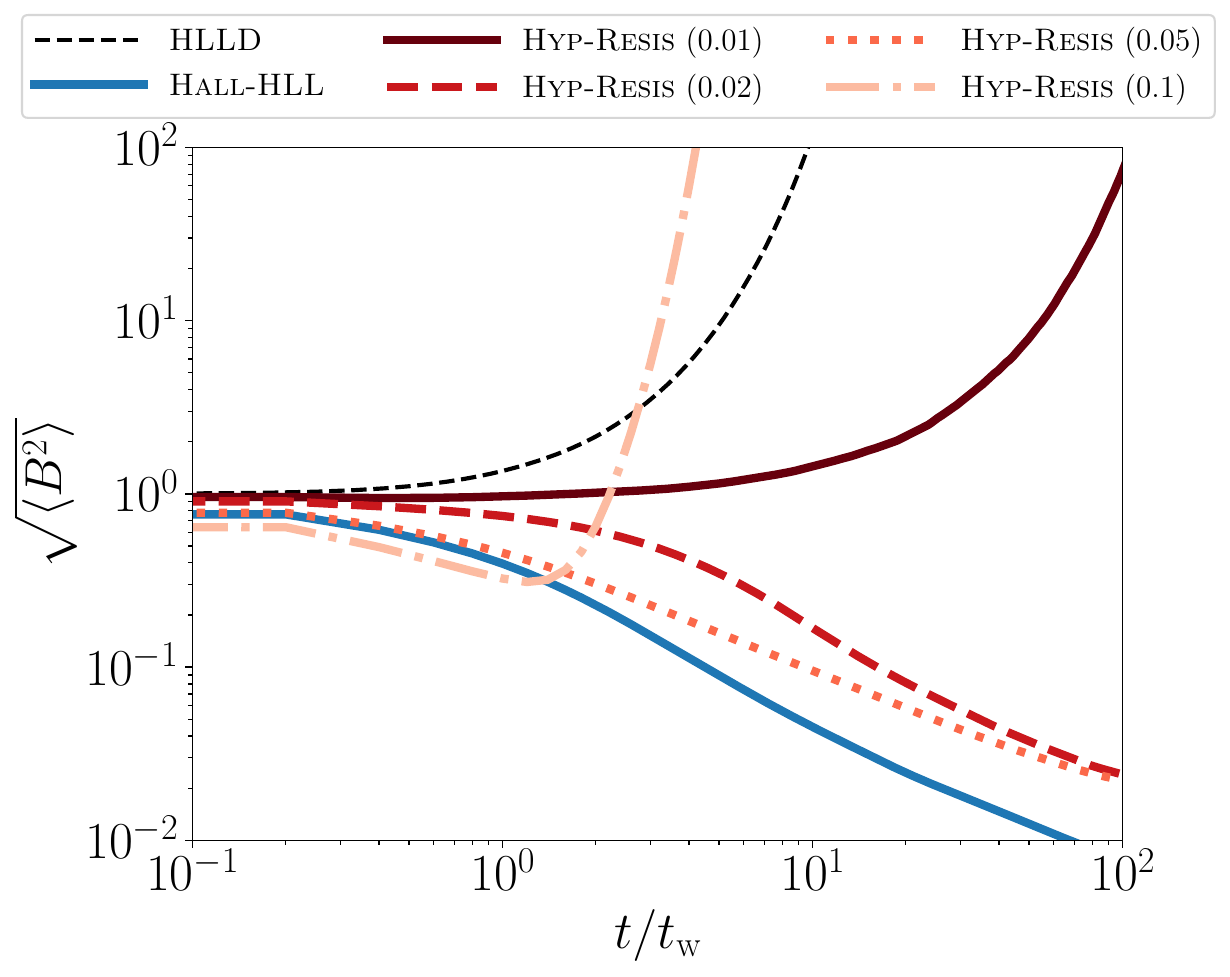}
    \caption{
    Time evolution of the magnetic field fluctuations.
    The results obtained using various methods ({\sc HLLD}, {\sc Hall-HLL}, 
    {\sc Hyp-Resis} with $C_\mathrm{hyp}=0.01$, $0.02$, $0.05$, $0.1$) are shown. 
    }
    \label{fig:dB_timeevo}
\end{figure}

\subsubsection{Comparison Between Different Implementations}\label{sec:turb_comp}

We compare the results from the four implementations listed in Table \ref{tab:methods}.
Here, we set $\etaH=10^4$ and $L=4$.
The results for {\sc Hall-HLLmod} are omitted in this section,
as they are nearly identical to those of {\sc Hall-HLL} 
because the gas remains nearly static in this test.

Comparison of the results obtained using {\sc HLLD}, 
{\sc Hall-HLL}, and {\sc Hyp-Resis} is 
shown in Figure \ref{fig:dB_timeevo}.
For {\sc Hyp-Resis}, 
the runs with $0.01<C_\mathrm{hyp}<0.1$ produce stable results, whereas 
those with $C_\mathrm{hyp}=0.01$ and $0.1$ show numerical instabilities.
This is roughly consistent with the requirement 
that $C_\mathrm{hyp}$ should be greater than $C_\mathrm{hyp,min}\sim 0.03$ and less than $C_\mathrm{hyp,max}$
(Section \ref{sec:Chypmin}). 
{\sc Hall-HLL} is also stable, but it reduces $\langle\sqrt{\delta \bm{B}^2}\rangle$ 
faster than {\sc Hyp-Resis}.

\begin{figure}[htpb]
    \centering
    \includegraphics[width=6cm]{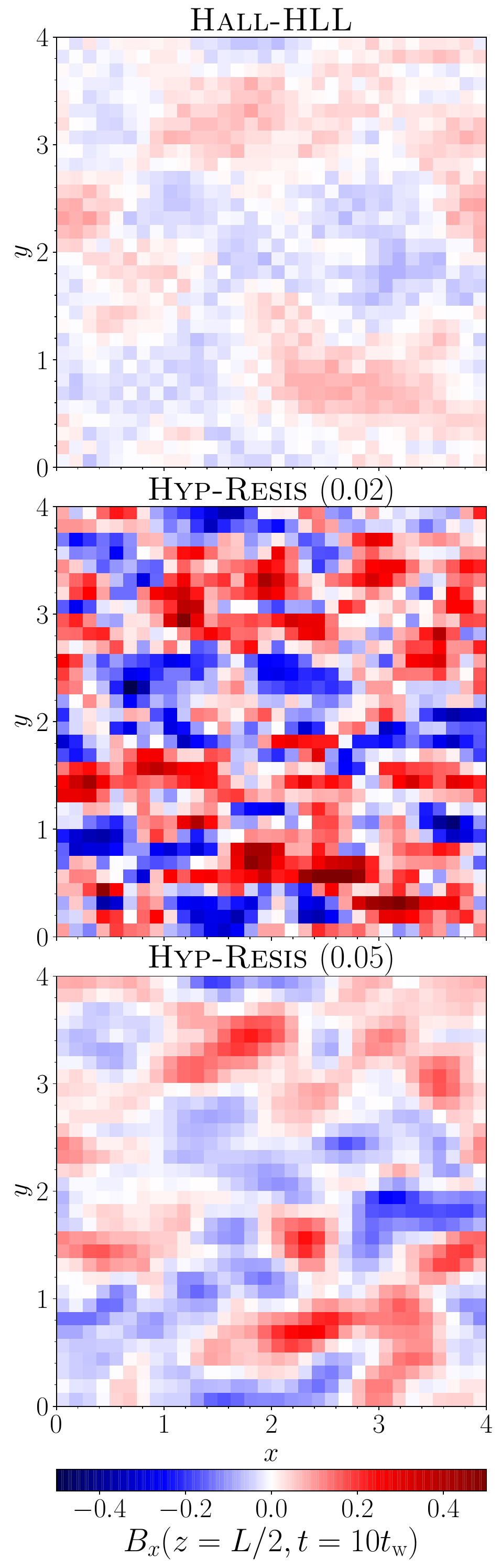}
    \caption{
    The $B_x$ slice at $z=2$ for 
    {\sc Hall-HLL}  and 
    {\sc Hyp-Resis} ($C_\mathrm{hyp}=0.02$ and $0.05$).
    The maps are taken at $t=10t_\mathrm{w}$.
    }
    \label{fig:turb_slice}
\end{figure}

In order to examine the magnetic field structure of the stable results, we present
the $B_x$ slice map at $z=L/2$ in Figure \ref{fig:turb_slice}.
Firstly, the results with {\sc Hyp-Resis} are analyzed.
Grid-scale fluctuations, which persist for $C_\mathrm{hyp}=0.02$, disappear 
in the $B_x$ maps for $C_\mathrm{hyp}=0.05$.

{\sc Hall-HLL} exhibits significantly distinct
features compared to {\sc Hyp-Resis}, as shown 
in Figure \ref{fig:turb_slice}.
At $t=10t_\mathrm{w}$,
grid-scale disturbances remain in {\sc Hall-HLL}, whereas
large-scale fluctuations are less pronounced in {\sc Hall-HLL} 
compared to {\sc Hyp-Resis} with $C_\mathrm{hyp}=0.05$.
This is due to the fact that {\sc Hall-HLL} utilizes the cell-centered transverse magnetic fields rather than 
the face-centered magnetic fields when computing the numerical fluxes (Section \ref{sec:implementation}).
Since cell-centered magnetic fields are derived by 
using a simple arithmetic average of face-centered magnetic fields
(Equation (\ref{face_to_center})), 
grid-scale disturbances of the face-centered magnetic field 
are significantly reduced during the conversion from the face-centered $\bm{B}$ to the cell-centered $\bm{B}$.
By contrast, the face-centered magnetic fields are used to compute the edge-centered electric field due to 
the hyper resisitivity.
Thus, {\sc Hall-HLL} is less effective in reducing pre-existing grid-scale perturbations in $\bm{B}$ than 
{\sc Hyp-Resis}, which directly utilizes the face-centered $\bm{B}$.

\subsection{Density-shear Instability}\label{sec:denshear}

In this section, 
we investigate the density-shear instability as a numerical experiment involving variable $\etaH$.
This instability occurs in situations where both the unperturbed magnetic field
$\bm{B}_0$ and the electron number density have 
steep gradients perpendicular to $\bm{B}_0$ \citep{Wood2014PhPl...21e2110W}.
The Hall effect coefficient $\etaH$ is given by 
$\etaH = cB/(4\pi en_e)$, where $c$ is the speed of light, 
$B$ is the magnetic field strength, 
$e$ is the elementary electric charge, and $n_e$ 
is the electron number density.

\citet{Gourgouliatos2015MNRAS.453L..93G} conducted 
numerical simulations of the density-shear instability.
We solve the full set of the Hall-MHD equations 
(Equations (\ref{eoc})-(\ref{induc})), whereas
they considered the induction equation taking into account only the Hall electric field 
using a pseudo-spectral code.
However, the results are expected to be consistent with those of \citet{Gourgouliatos2015MNRAS.453L..93G}
because the gas remains nearly static as 
the magnetic field evolves,
similar to the turbulent magnetic field test (Section \ref{sec:whitenoise}).

Following \citet{Gourgouliatos2015MNRAS.453L..93G}, 
we assume an unperturbed plane-parallel magnetic field, localized around $y=0$ with a characteristic length scale of $a$
as follows: 
\begin{equation}
  B_x = B_0 \left\{ \exp\left( - \frac{y^2}{a^2}\right) + \epsilon_B \right\},
  B_y=B_z=0,
  \label{unpB}
\end{equation}
where $\epsilon_B=10^{-2}$ represents the magnetic field floor.
The electron number density has a similar 
functional form to $B_x$:
\begin{equation}
  n_e = n_0 \left\{ \exp\left( - \frac{y^2}{a^2}\right) + \epsilon_n \right\},
  \label{ne}
\end{equation}
where $\epsilon_n=10^{-2}$ is a floor bound for $n_e$.
We consider a fully ionized gas in which 
the density is proportional to $n_e$ and 
is $\rho_0=1$ at $y=0$.
The initial gas pressure is set so that the total pressure 
$P+B_x^2/8\pi$ is spatially constant. 
The results are insensitive to the total pressure 
because the gas remains nearly static throughout the evolution.
In this study, we set $8\pi P/B_x^2 = 0.1$ for $y=0$ and $B_0=1$.

From Equation (\ref{ne}), the Hall effect coefficient is expressed as
\begin{equation}
    \etaH = \eta_\mathrm{H0}\frac{|\bm{B}|/B_0}{\rho/\rho_0},
\end{equation}
where $\eta_\mathrm{H0}=cB_0/(4\pi e n_0)$ is the reference 
Hall effect coefficient and is assigned a value of 100.

\citet{Wood2014PhPl...21e2110W} demonstrated through linear analysis that 
the growth rate reaches its maximum when the perturbation wavenumber vector is aligned 
to the $x$-axis.
The dispersion relation is given by 
\begin{equation}
    \sigma(k) = \frac{\eta_\mathrm{H0}}{a^2}\sqrt{(ka)^2\{2-(ka)^2\}},
\end{equation}
where $\sigma$ represents the growth rate and $k$ denotes 
the wavenumber along the $x$-axis
\citep{Wood2014PhPl...21e2110W}.
The maximum growth rate $\sigma_\mathrm{max}=\eta_{H0}/a^2$ is obtained at $k = a^{-1}$.

The fastest growing mode is applied to $B_y$ as follows:
\begin{eqnarray}
    B_y = \delta B\cos\left(x/a\right),
\end{eqnarray}
where $\delta B=10^{-4}B_0$ is the initial amplitude.
No perturbations are introduced for the other variables.

The simulations are performed in two dimensions.
The box size in the ($x$,$y$) plane spans 
$-3\pi a\le x,y\le 3\pi a$ and is discretized 
by $128^2$ cells, where the characteristic scale is $a=0.1$.
The width $a$ is resolved by $\sim 14$ cells.
We evaluate the four different implementations: {\sc HLLD}, {\sc Hall-HLL}, {\sc Hall-HLLmod}, and {\sc Hyp-Resis}. 
To investigate an appropriate value of $C_\mathrm{hyp}$, 
we test $C_\mathrm{hyp} = 0.01,0.02,0.05,0.1$.

\begin{figure}[htpb]
    \centering
    \includegraphics[width=8cm]{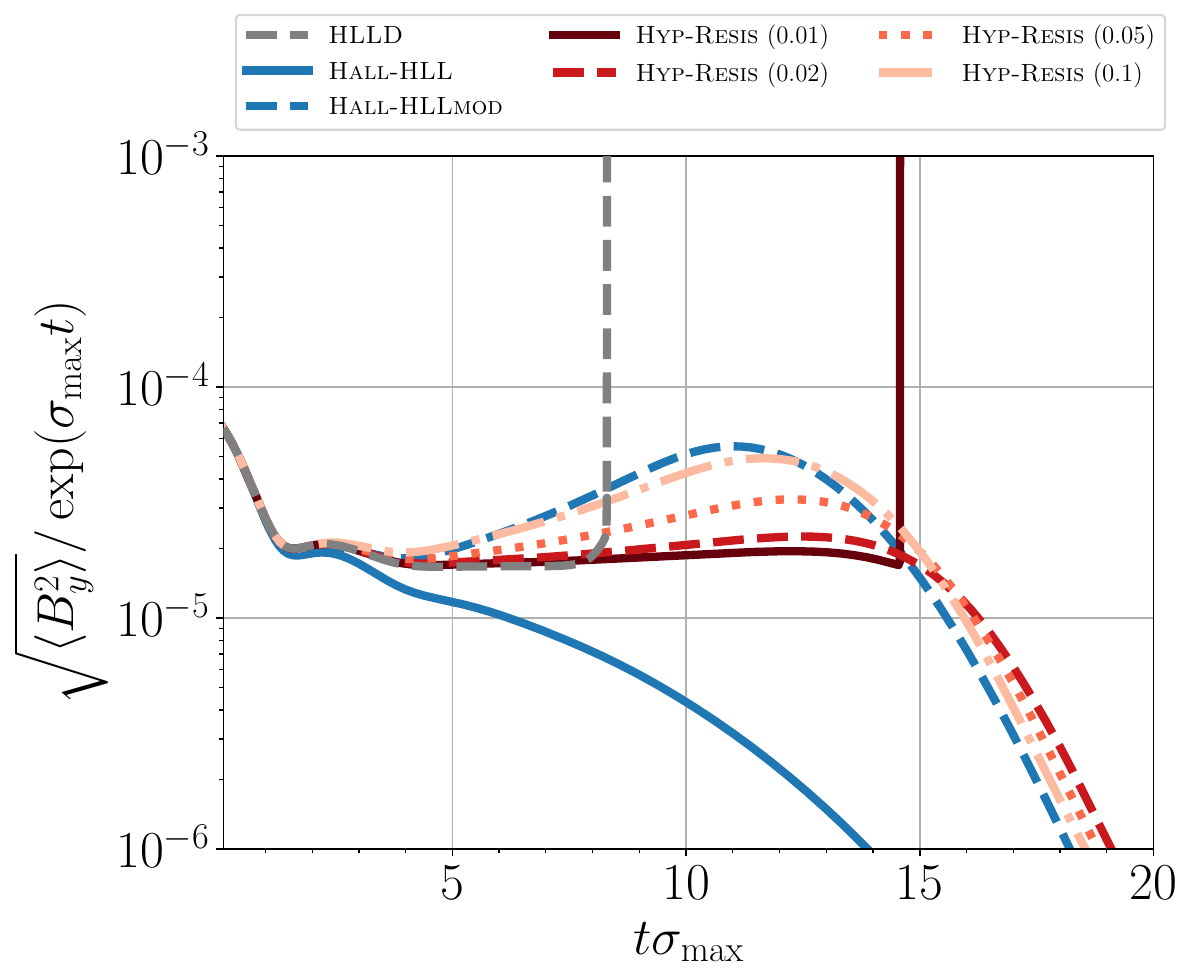}
    \caption{
    Time evolution of $\sqrt{\langle B_y^2\rangle}$ divided by $\exp(\sigma_\mathrm{max}t)$ 
    for the four different implementations ({\sc HLLD}, {\sc Hall-HLL}, {\sc Hall-HLLmod}, 
    and {\sc Hyp-Resis} with $C_\mathrm{hyp}=0.01,0.02,0.05,$ and $0.1$).
    The horizontal axis is normalized by $\sigma_\mathrm{max}^{-1}$.
    }
    \label{fig:denshear_dB_timeevo}
\end{figure}

\subsubsection{Linear Growth}\label{sec:denshear_linear}

First, the early-time evolution of $\langle B_y^2\rangle^{1/2}$ 
is compared to the theoretical predictions from the linear analysis in Figure \ref{fig:denshear_dB_timeevo}.
For {\sc HLLD}, the evolution of $\langle B_y^2\rangle^{1/2}$
agrees with the expected trend 
$\exp(\sigma_\mathrm{max}t)$ until $t\sigma_\mathrm{max}\sim 8$, after which 
the numerical instability rapidly increases.

For {\sc Hall-HLL}, $\langle B_y^2\rangle^{1/2}$ 
increases significantly slower than $\exp(\sigma_\mathrm{max}t)$
due to large dissipation introduced in both 
the velocity and magnetic fields.
The initially concentrated distributions of $\rho$ and $B_x$ along the $y$-axis (Equations (\ref{unpB}) and (\ref{ne}))
become significantly diffused.

Next, we analyze the {\sc Hyp-Resis} runs.
When $C_\mathrm{hyp}=0.01$, the growth rate agrees with
$\sigma_\mathrm{max}$, but as 
$C_\mathrm{hyp}$ increases from 0.01, the growth rate departs from $\sigma_\mathrm{max}$.
Counterintuitively, increasing $C_\mathrm{hyp}$ leads to 
an increased growth rate.
This occurs because artificial diffusion is introduced 
exclusively in the magnetic field.
As a result, the profiles of $B_x$ undergo diffusion,
while those of $\rho$ remain nearly unchanged.
This can be interpreted as setting a non-perturbed state where the width of the $B_x$ profile ($a_{B}$)
is slightly greater than that of the $\rho$ profile ($a_\rho$).
The growth rate increases with $a_B$ at a fixed $a_\rho$ \citep{Gourgouliatos2015MNRAS.453L..93G}.

{\sc Hyp-Resis} with lower $C_\mathrm{hyp}$ may experience 
numerical instabilities.
Figure \ref{fig:denshear_dB_timeevo} indicates that 
the numerical instability arises for {\sc Hyp-Resis} 
with $C_\mathrm{hyp}=0.01$ around $t\sim 14.5$.
The {\sc Hyp-Resis} runs with $C_\mathrm{hyp}>0.01$ 
appear to provide numerically stable results.

{\sc Hall-HLLmod} produces more accurate results than {\sc Hall-HLL}.
The time evolution of $\langle B_y^2\rangle^{1/2}$ is close to that of 
{\sc Hyp-Resis} with $C_\mathrm{hyp}=0.1$.
This occurs because the density profile diffusion 
is suppressed in {\sc Hall-HLLmod}.

\begin{figure}[htpb]
    \centering
    \includegraphics[width=8.5cm]{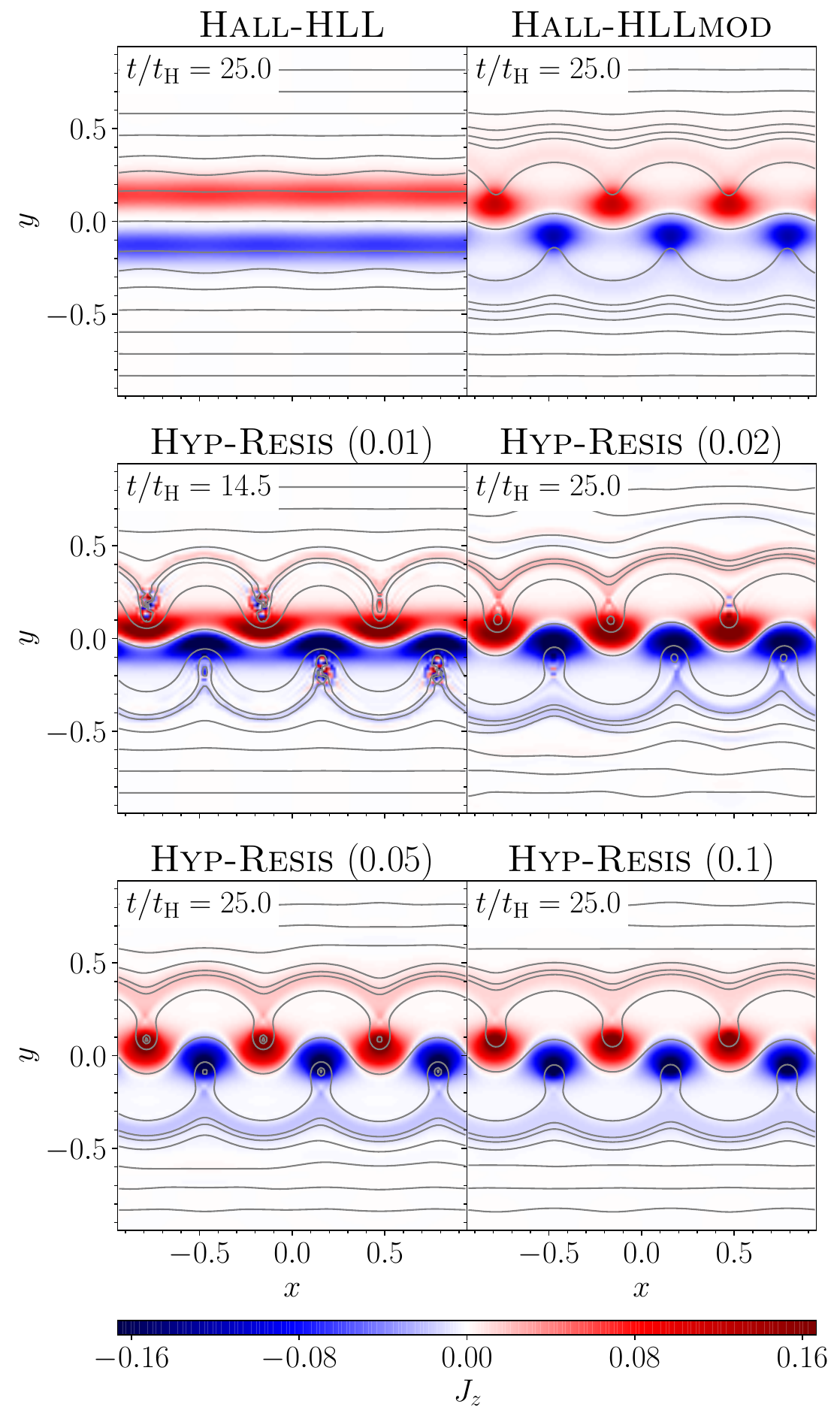}
    \caption{
    Color maps of the current density along the $z$-axis, $J_z$, for 
    four different implementations ({\sc HLLD}, {\sc Hall-HLL}, {\sc Hall-HLLmod}, 
    and {\sc Hyp-Resis} with $C_\mathrm{hyp}=0.01,0.02,0.05,$ and $0.1$).
    The gray lines indicate the field lines in the $x$-$y$ plane.
    The snapshots are taken at $t/t_\mathrm{H}=25$. 
    For {\sc Hyp-Resis} with $C_\mathrm{hyp}$=0.01, the $J_z$ map is shown 
    at the time when the numerical instability occurs.
    }
    \label{fig:denshear_slice}
\end{figure}

\subsubsection{Nonlinear Evolution}

Next, we investigate the nonlinear evolution of the density-shear instability.
The snapshots at $t=25$ are displayed in Figure \ref{fig:denshear_slice}.
Corrugations in the magnetic field around $y=0$ develop over time.
When magnetic fields deform sufficiently, 
magnetic reconnection is triggered around pinched magnetic fields.

As discussed in Section \ref{sec:denshear_linear}, 
the numerical instability arises at $t=14.5$ in the 
{\sc Hyp-Resis} case with $C_\mathrm{hyp}=0.01$, particularly in the reconnection 
regions where $J_z$ exhibits significant fluctuations.
In the $J_z$ map of {\sc Hyp-Resis} with $C_\mathrm{hyp}=0.02$,
numerical wiggles are also seen around $(x,y)\sim (0.5,0.2)$ and $(-0.5,-0.2)$.
This indicates that $C_\mathrm{hyp}=0.02$ leads to numerical fluctuations due to insufficient 
dissipation.
This is consistent with the fact that $C_\mathrm{hyp}=0.02$ is less than $C_\mathrm{hyp,min}\sim 0.03$ 
(Equation (\ref{Chypmin})).
No numerical oscillations are visible in the map of {\sc Hyp-Resis} with $C_\mathrm{hyp}\ge 0.05$.

Compared to {\sc Hyp-Resis}, {\sc Hall-HLL} produces 
significantly more diffused results.
The field lines are almost straight, indicating 
that the density shear instability is almost suppressed.
While {\sc Hall-HLLmod} significantly improves the dissipative distribution of $J_z$,
Figure \ref{fig:denshear_slice} reveals that 
$J_z$ in {\sc Hall-HLLmod} remains more diffusive than {\sc Hyp-Resis} with $C_\mathrm{hyp}=0.1$.

\begin{figure*}[htpb]
    \centering
    \includegraphics[width=0.9\textwidth]{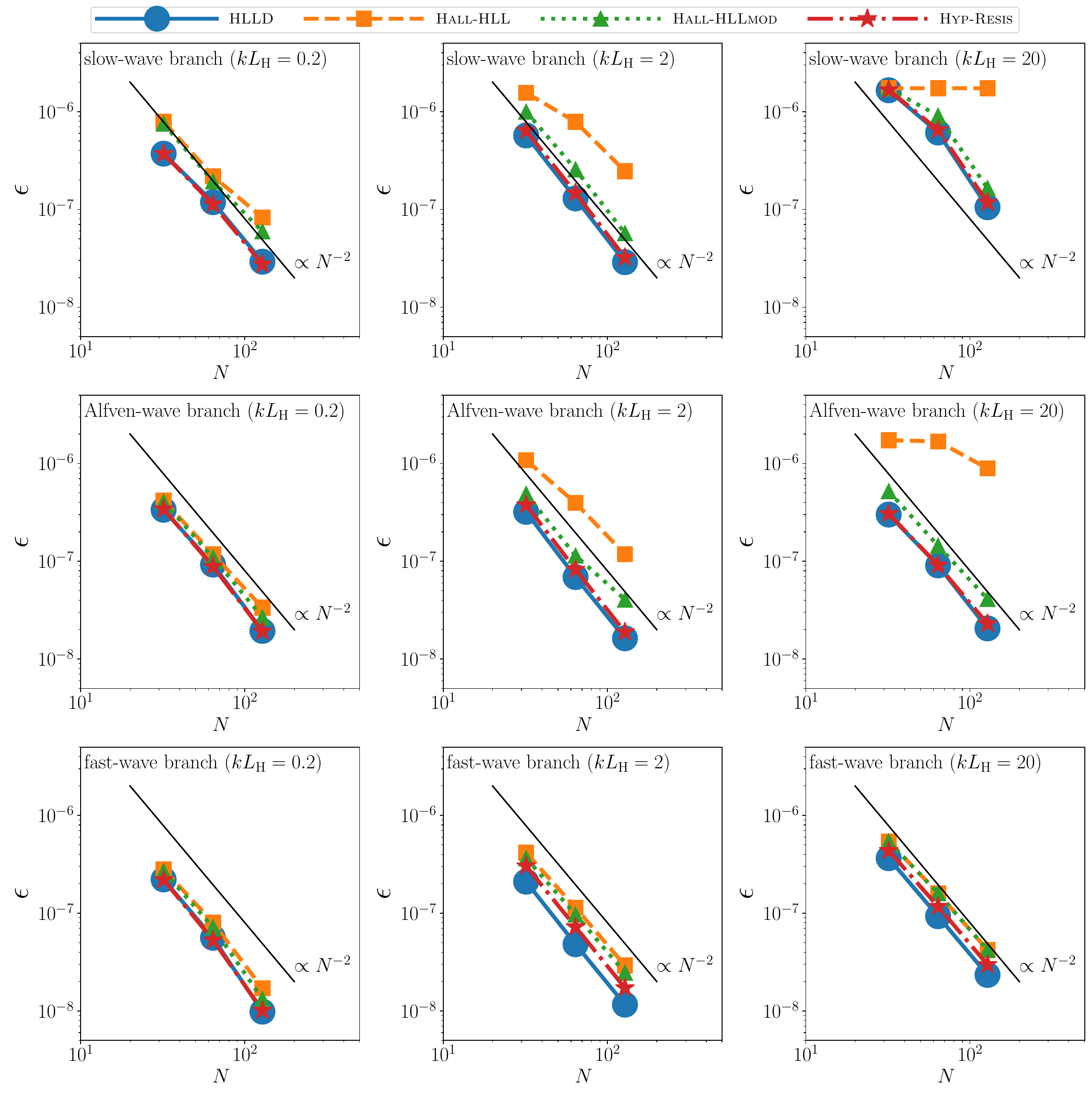}
    \caption{
    Results of the convergence test ($\tcs=1/2$ and $\theta=\pi/4$) for {\sc HLLD},
    {\sc Hall-HLL}, 
    {\sc Hall-HLLmod}, and {\sc Hyp-Resis} with $C_\mathrm{hyp}=0.05$. 
    The left, middle, and right columns correspond to the results for
    $kL_\mathrm{H}=0.2$, $k\LH=2$, and $k\LH=20$, respectively.
    The top, middle, and bottom rows correspond to 
    the slow-wave branch, Alfv\'en-wave branch, and 
    fast-wave branch, respectively.
    For reference, the black solid lines proportional to $N^{-2}$ are plotted.
   }
    \label{fig:convergence}
\end{figure*}

\subsection{Linear Wave Convergence Test}\label{sec:linearwaves}

In Section \ref{sec:linearanalysis}, 
the physical properties of the fast-wave branch, Alfv\'en-wave branch, and slow-wave branch 
were discussed.
In this section, we conduct the convergence tests of linear waves in both uniform and 
static mesh refinement (SMR) grids. 

The setup is as follows:
we consider propagation of linear waves at an inclination relative to the grid cells.
The numerical setup follows that of Section \ref{sec:rk3_hall}, except that 
all types of the linear waves are considered in this section.
In the coordinate system $\bm{\xi}=(\xi,\eta,\zeta)$ defined as ${\cal R}\cdot \bm{x}$,
The perturbation vector is defined as
\begin{equation}
    \delta \bm{Q} = \left( c_\mathrm{s}\frac{\delta \rho}{\rho_0}, \delta v_\xi, \delta v_\eta, \delta v_\zeta, 
    \frac{\delta B_\xi}{\sqrt{4\pi\rho_0}},
    \frac{\delta B_\eta}{\sqrt{4\pi\rho_0}},
    \frac{\delta B_\zeta}{\sqrt{4\pi\rho_0}}
    \right), 
\end{equation}
and has a spatial dependence of $e^{i\bm{k}\cdot\bm{\xi}}$.
The initial perturbation amplitude is set to $|\delta \bm{Q}|=10^{-6}c_\mathrm{A}$.

The simulation box spans $0\le x \le L$, $0\le y,z\le L/2$ and 
is divided by $N\times (N/2)\times (N/2)$ cells, where $L=6\pi/k$.
For a given $k$, the eigenfunctions of the 
three branches are considered in
the coordinates $(x,y,z)$ as the initial conditions.
The volume-weighted $L_2$ norm is measured at 
$t=2\pi/\omega(k)$, is defined as follows:
\begin{equation}
    \epsilon = \sqrt{
    \frac{\displaystyle\sum_{n} \sum_{i,j,k}
    \left(\delta Q_{n,i,j,k} 
    - \delta Q_{n,\mathrm{exact}}(\bm{x}_{i,j,k})\right)^2\Delta V_{i,j,k}}
    {\displaystyle \sum_{i,j,k}\Delta V_{i,j,k}}},
\end{equation}
where $\bm{x}_{i,j,k}=(x_i,y_j,z_k)$, $\delta Q_{n,\mathrm{exact}}$ represents 
the exact solution of the $n$-th component of $\delta\bm{Q}$
at $t=2\pi/\omega(k)$, and $\Delta V_{i,j,k}$ denotes the volume of 
the cell centered at $\bm{x}_{i,j,k}$.
We consider the case where $\tcs=1/2$ and $\theta=\pi/4$, with 
the dispersion relation shown in Figure \ref{fig:disp_hallmhd}.

We compare the results obtained by 
the four different methods ({\sc HLLD}, {\sc Hall-HLL}, 
{\sc Hall-HLLmod}, {\sc Hyp-Resis}).
For the {\sc Hyp-Resis} runs, $C_\mathrm{hyp}$ is fixed to $0.05$ 
because this value provides the minimum dissipation required 
to eliminate the numerical fluctuations (Sections \ref{sec:whitenoise} and \ref{sec:denshear}).

\subsubsection{Uniform Grids}\label{sec:uniformgrid}

For the uniform grids, 
we consider three different wavelengths 
($k\LH=0.2$, 2, 20), spanning from the ideal MHD regime to 
the Hall-dominated regime (Figure \ref{fig:disp_hallmhd}).
The convergence test is performed by changing $N$ ($N=32$, 64, and 128) using 
$3\times 3\times 4$ combination of the three branches, 
three different wavelengths, and 
the four different methods ({\sc HLLD}, {\sc Hall-HLL}, 
{\sc Hall-HLLmod}, {\sc Hyp-Resis}).
The results are summarized in Figure \ref{fig:convergence}.

\begin{table}[htpb]
    \centering
    \begin{tabular}{|c|c|c|c|}
    \hline
    $k\LH$ & $N=32$ & $N=64$ & $N=128$  \\
    \hline
    \hline
    0.2 & 1.06 & 1.25 & 1.69 \\
    \hline
    2 & 3.28 & 6.20 & 12.2 \\
    \hline
    20 & 16.9 & 30.8 & l21 \\
    \hline
    \end{tabular}
    \caption{The values of $c_\mathrm{w}(k_\mathrm{max}=\Delta x^{-1})$ divided by 
    $\sqrt{c_\mathrm{s}^2+c_\mathrm{A}^2}$, which 
    is the maximum value of $c_\mathrm{f}$}
    \label{tab:cw_cf}
\end{table}

Both {\sc Hyp-Resis} and {\sc HLLD} exhibit second-order convergence
across all the branches and wavelengths.
The errors for {\sc Hyp-Resis} and {\sc HLLD} 
are nearly identical, indicating that the hyper-resistivity 
with $C_\mathrm{hyp}=0.05$ does not significantly dampen linear waves.


Next, the results of {\sc Hall-HLL} are compared with those of {\sc HLLD}.
The importance of the Hall effect in the signal speed in the {\sc Hall-HLL} flux
is illustrated in Table \ref{tab:cw_cf} that shows that 
the values of $c_\mathrm{w}(\Delta x^{-1})/\sqrt{\cA^2+c_\mathrm{s}^2}$
increase with increasing $k\LH$ and $N$.

For $k\LH=0.2$, the Hall effect does not significantly affect the signal speed of the {\sc Hall-HLL} flux 
since $c_\mathrm{w}(\Delta x^{-1}) \sim \sqrt{c_\mathrm{s}^2+\cA^2}$.
{\sc Hall-HLL} exhibits second-order convergence, 
similar to {\sc Hyp-Resis} and {\sc HLLD} for all branches.
In the slow-wave branch, the error $\epsilon$ in {\sc Hall-HLL} is approximately twice as large as 
that in {\sc HLLD}, whereas the errors are comparable for the 
fast-wave and Alfv\'en-wave branches.
This discrepancy arises because the numerical dissipation in 
{\sc Hall-HLL} is determined by the phase speed of fast waves.

At larger wavenumbers, $k\LH=2$ and $20$, 
{\cal Hall-HLL} behaves differently from {\sc Hyp-Resis} and {\sc HLLD} because 
the signal speeds in the {\sc Hall-HLL} solver are determined by whistler waves.
For the fast-wave branch,
{\sc Hall-HLL} shows second-order convergence, and 
the errors $\epsilon$ are comparable to those of {\sc Hyp-Resis} and {\sc HLLD} because the fast-wave branch 
corresponds to whistler waves.
However, for the Alfv\'en-wave and slow-wave branches especially at the largest wavenumber $k\LH=20$, 
the errors in {\sc Hall-HLL} are much larger than those of {\sc Hyp-Resis} and {\sc HLLD} and 
decrease with $N$ at a slower rate than second-order 
convergence, at least in the range $N\le 128$.
This occurs because the {\sc Hall-HLL} solver incorporates the phase speed of the whistler waves, 
leading to significant numerical dissipation that dampens the linear waves
with  phase speeds much smaller than $c_\mathrm{w}$ (Table \ref{tab:cw_cf}).
Both sound and ion-cyclotron waves are significantly
dampened at $k\LH= 2$ and $20$ as shown 
in Figure \ref{fig:convergence}.


Figure \ref{fig:convergence} shows that
{\sc Hall-HLLmod} significantly improves the dissipative properties of {\sc Hall-HLL} for the Alfv\'en-wave and 
slow-wave branches, and restores second-order convergence.
However, the error $\epsilon$ in 
{\sc Hall-HLLmod} is still larger than those in {\sc Hyp-Resis}.
This indicates that {\sc Hyp-Resis} can capture all MHD linear waves more accurately than both
{\sc Hall-HLL} and {\sc Hall-HLLmod}.

We compare the performance of {\sc HLLD}, {\sc Hall-HLLmod},
and {\sc Hyp-Resis} at the highest resolution, $N=128$.
These calculations are conducted on a single HPE Cray XD2000 node with
dual Intel Xeon CPU Max 9480 processors.
The calculation using {\sc Hyp-Resis} is only 3\% slower than those using 
{\sc HLLD}, 
indicating that the computational cost of the hyper-resistivity term is negligible.
The computational speed of {\sc Hall-HLLmod} is about 
11\% (12\%) faster than that of {\sc HLLD} ({\sc Hyp-Resis})
since the Riemann solver of {\sc Hall-HLLmod}
is computationally cheaper than that of the HLLD solver.
However, to achieve the desired accuracy in the simulations, 
the total number of cells required with 
{\sc Hyp-Resis} can be reduced by a factor of $2^{3/2} \sim 2.8$ compared to {\sc Hall-HLL}.
In this argument, we use the fact that $\epsilon\propto N^{-2}$ 
and the errors of {\sc Hyp-Resis} 
are about half of those of {\sc Hall-HLL} (Figure \ref{fig:convergence}).

\subsubsection{Static Mesh Refinement Grids}\label{sec:whistlear_smr}

In this section, we examine whistler wave propagation 
in SMR grids to evaluate
the effect of mesh refinement on numerical stability and global convergence rates.
In this analysis, only the fast-wave branch with $k\LH=20$ is considered.

The root grid resolution is set as $N_\mathrm{root}\times(N_\mathrm{root}/2)\times(N_\mathrm{root}/2)$
Refined grids are introduced in the central region, 
$0.25L\le x\le 0.75L$ and $0.125L\le y,z\le 0.625L$, 
with cell sizes reduced to half of the root grid size.

\begin{figure}[htpb]
    \centering
    \includegraphics[width=8cm]{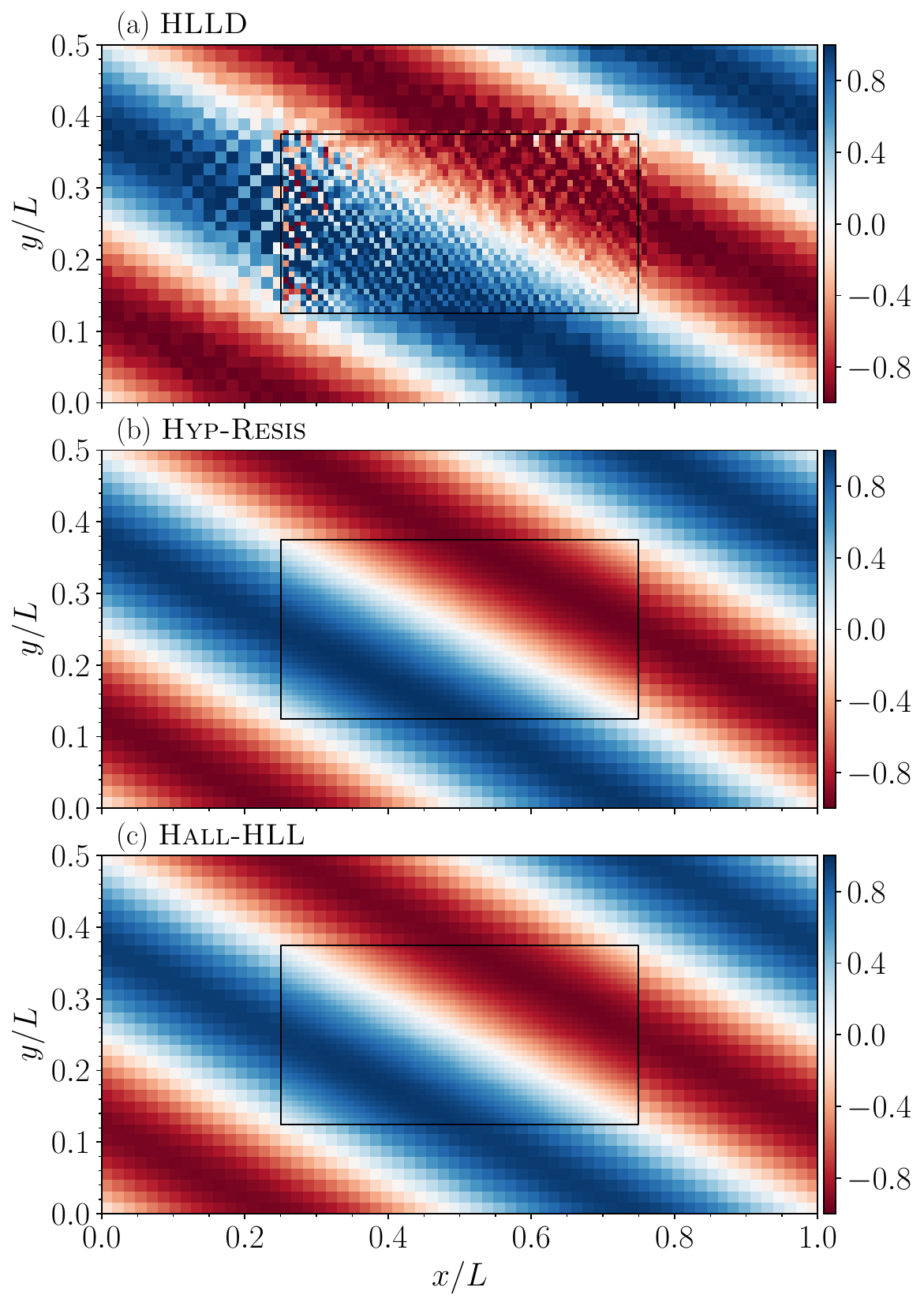}
    \caption{
    $B_\zeta$ maps at the plane 
    $z=L/4$ for (a) {\sc HLLD}, (b) {\sc Hyp-Resis} with 
    $C_\mathrm{hyp}=0.05$, and (c) {\sc Hall-HLL}.
    The snapshots are taken at $t=\pi/\omega(k)$.
    In each panel, 
    the rectangle encloses the refined region.
    }
    \label{fig:slice_smr}
\end{figure}

Figure \ref{fig:slice_smr} shows the $B_\zeta$ maps at $z=L/4$ for {\sc HLLD},
{\sc Hyp-Resis} with $C_\mathrm{hyp}=0.05$, and {\sc Hall-HLL}.
Unlike uniform grids, numerical instabilities arise in SMR grids.
Small-scale waves are excited near the level boundaries and 
grow over time.
Both {\sc Hyp-Resis} and {\sc Hall-HLL} produce stable results, as illustrated in 
Figures \ref{fig:slice_smr}b and \ref{fig:slice_smr}c.
A comparison of these panels reveals no significant
difference in $B_\zeta$ between {\sc Hyp-Resis} and {\sc Hall-HLL}.

\begin{figure}[htpb]
    \centering
    \includegraphics[width=8cm]{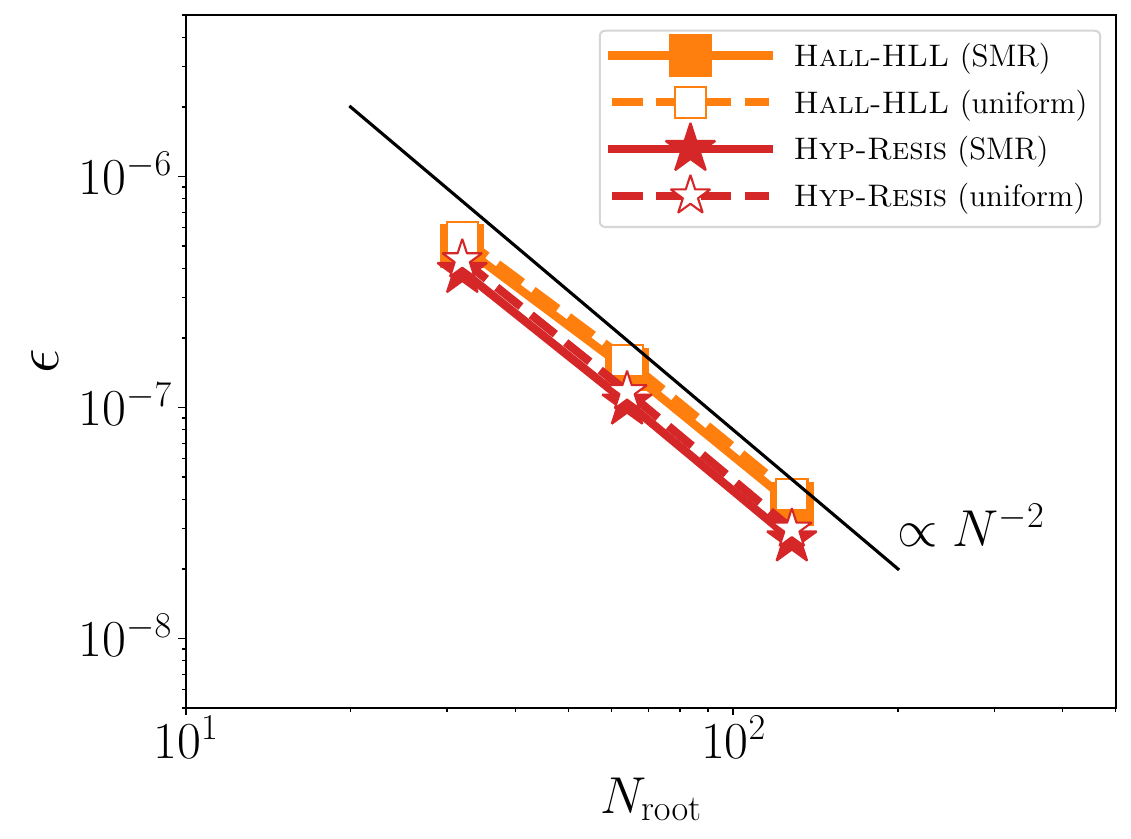}
    \caption{
    Convergent test of whistler waves with $\tilde{c}_\mathrm{s}=1/2$ and $k\LH=20$ for 
    {\cal Hall-HLL} and {\cal Hyp-Resis}.
    The L2 norms measured at $t=2\pi/\omega(k)$ 
    are plotted as a function of $N_\mathrm{root}$.
    For comparison, the results for the uniform grid are shown with dashed lines.
    }
    \label{fig:convergence_smr}
\end{figure}

Figure  \ref{fig:convergence_smr} presents  the global convergence rates.
Both {\sc Hall-HLL} and {\sc Hyp-Resis} demonstrate second-order global convergence.
As shown in Figure \ref{fig:convergence}, the errors $\epsilon$ are slightly smaller for {\sc Hyp-Resis} than 
for {\sc Hall-HLL}.
Furthermore, in each implementation, 
the SMR grid results in smaller errors compared to
the uniform grid, as expected.

\subsection{Kelvin-Helmholtz Instability}\label{sec:KH}

The Kelvin-Helmholtz (KH) instability in the presence of the Hall effect was investigated by 
\citet{TalwarKalra1967JPlPh...1..145T} and \citet{SenChou1968CaJPh..46.2557S}.
Their findings indicate that for super-Alfv\'enic shear flows, the Hall effect 
enhances the growth rate. 
\citet{Pandey2018MNRAS.476..344P} demonstrated that 
the Hall effect destabilizes 
sub-Alfv\'enic shear flows, which 
are stable under ideal MHD conditions
\citep{Chandrasekhar1961hhs..book.....C}. 

\subsubsection{Predictions from Linear Analyses}

Before presenting the numerical setup, 
we provide a brief overview of the growth rate derived by 
\citet{Pandey2018MNRAS.476..344P}.
The unperturbed state consists of 
a uniform gas with $v_x(y) = \machA c_\mathrm{A}$ 
for $y\ge 0$ and $v_x(y)=-\machA c_\mathrm{A}$ for $y<0$,
where $c_\mathrm{A}$ denotes the Alfv\'en speed of the 
unperturbed state, and $\machA$ is the Alfv\'en Mach number of the shear flow.
The magnetic field is uniform and aligned with the shear flow
along the $x$-axis.
By considering perturbations in the form $e^{\sigma t + ikx}$,
the following dispersion relation is obtained by applying 
the appropriate boundary conditions \citep{TalwarKalra1967JPlPh...1..145T,SenChou1968CaJPh..46.2557S,Pandey2018MNRAS.476..344P},
\begin{eqnarray}
&& 
(2 + \tilde{\sigma}_2^2 + \tilde{\sigma}_1^2) 
\left\{
Q_1\tilde{\sigma}_1(\tilde{\sigma}_2^2 + 1)
+ Q_2\tilde{\sigma}_2 (\tilde{\sigma}_1^2 + 1)
\right\}\nonumber\\
&& \hspace{1cm} + kL_\mathrm{H}(\tilde{\sigma}_2^2 - \tilde{\sigma}_1^2)^2 =0,
\label{dispkh}
\end{eqnarray}
where $\tilde{\sigma}_1 =\sigma/(k\cA) + i \machA$, 
$\tilde{\sigma}_2 =\sigma/(k\cA) - i \machA$, 
\begin{equation}
    Q_j = \sqrt{
    (\tilde{\sigma}_j kL_\mathrm{H})^2 + (\tilde{\sigma}_j^2 + 1)^2
    }.
\end{equation}

\begin{figure}[htpb]
    \centering
    \includegraphics[width=8cm]{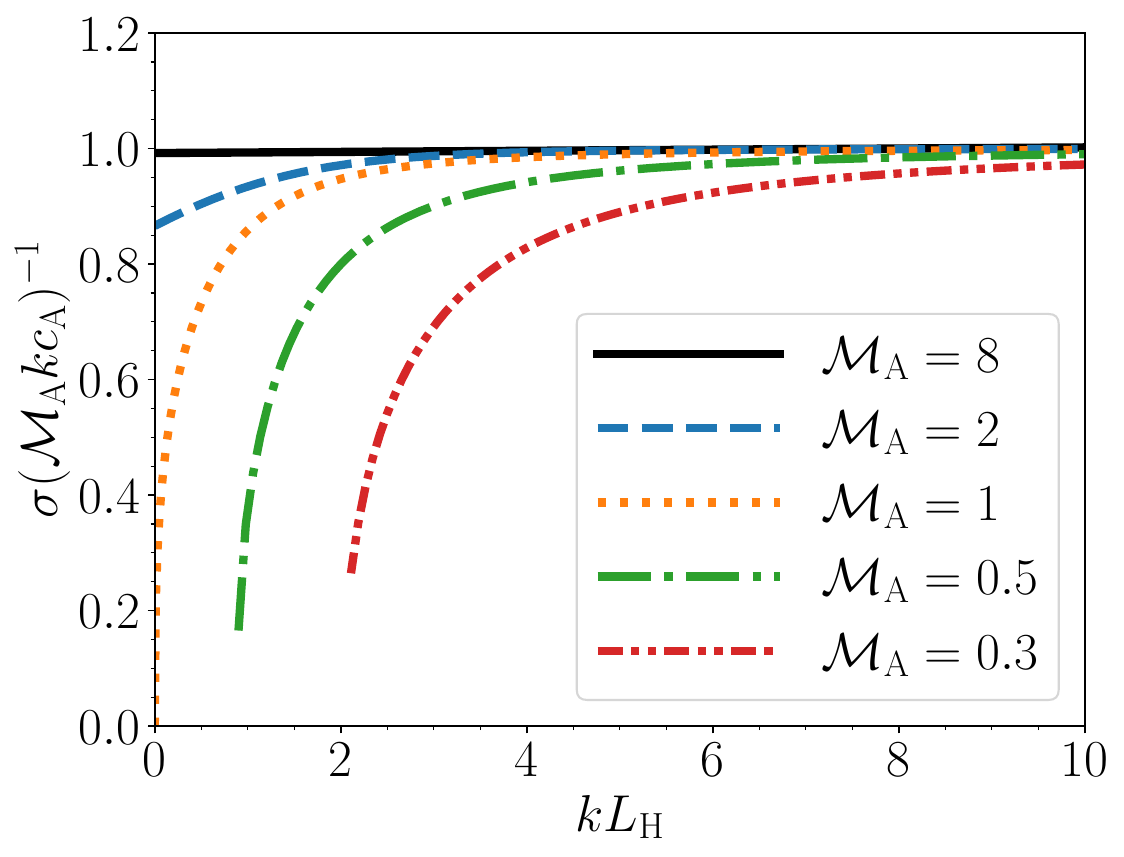}
    \caption{
    Growth rate of the KH instability as a function of $kL_\mathrm{H}$ 
    for $\machA=8$, 2, 1, 0.5, and 0.3.
    }
    \label{fig:disp_kh}
\end{figure}

Figure \ref{fig:disp_kh} shows the growth rate of 
the purely growing mode, which develops without oscillations, as
a function of $kL_\mathrm{H}$ for various values of $\machA$.

For the Hall-dominated limit ($kL_\mathrm{H}\gg 1$),
$Q_j$ can be approximated as $\tilde{\sigma_j} k\LH$.
Consequently, Equation (\ref{dispkh}) becomes independent 
of $k\LH$ and simplifies to:
\begin{eqnarray}
&& 
\left\{
\left(\frac{\omega}{k\cA}\right)^2 
+ (\machA - 1)^2
\right\}
\left\{
\left(\frac{\omega}{k\cA}\right)^2 
+ (\machA + 1)^2
\right\} \nonumber \\
&& \times 
\left( 
\frac{\omega}{k\cA} - \machA
\right)
\left( 
\frac{\omega}{k\cA} + \machA
\right) 
=0.
\label{dispkh_hall}
\end{eqnarray}
Equation (\ref{dispkh_hall}) has one purely growing mode given by 
\begin{equation}
    \sigma_\mathrm{hall} = kc_\mathrm{A}\machA.
    \label{sigmahall}
\end{equation}
This behavior differs significantly from the ideal MHD case, 
where sub-Alfv\'enic shear flows are stabilized by the Lorentz force.

For small values of $kL_\mathrm{H}$, 
the properties of $\sigma$ differ between super- and sub-Alfv\'enic cases.
In the super-Alfv\'enic regime, 
at $kL_\mathrm{H}=0$,
the growth rate is identical to that in ideal MHD,
\begin{eqnarray}
    \sigma_\mathrm{ideal} = kc_\mathrm{A}\sqrt{\machA^2-1}
\end{eqnarray}
\citep{Chandrasekhar1961hhs..book.....C}.
As $k L_\mathrm{H}$ increases, $\sigma$ increases and 
asymptotically approaches $\sigma_\mathrm{hall}$.
Conversely, in the sub-Alfv\'enic regime, 
Figure \ref{fig:disp_kh} shows that
purely growing modes exist only when $k L_\mathrm{H}$ 
exceeds a critical 
value, which is larger for smaller values of $\machA$ \citep{Pandey2018MNRAS.476..344P}.

\subsubsection{Numerical Setting}

The two-dimensional computational domain is defined as
$|x|\le 1/2$ and $|y|\le 1$, and 
is discretized into $256\times 512$ cells.
A uniform gas with the density $\rho_0$ 
and the pressure $P_0$ is considered.
The initial unperturbed gas flow and magnetic field 
are aligned with the $x$-axis. 
To impose periodic boundary conditions in all directions, 
two initial discontinuities are set at $y=\pm 1/2$.
Instead of considering perfect discontinuities in the $v_x$ profile, which 
is consistent with the unperturbed state in the linear analysis presented above, 
a smoothed $v_x$ profile is employed to achieve convergence of results \citep{Lecoanet2016MNRAS.455.4274L}.
The unperturbed gas flow is given by 
\begin{eqnarray}
    v_x(y) &=& \machA \cA \nonumber \\
    & \times& \left\{
    \tanh\left( \frac{y+1/2}{w_0} \right)
    - \tanh\left( \frac{y-1/2}{w_0} \right)
    -1
    \right\},
\end{eqnarray}
where $w_0=0.1$ is a parameter that controls the smoothness of the profile, 
$\cA = B_0/\sqrt{4\pi\rho_0}$ is the initial Alfv\'en speed.
To track the time evolution of the initial discontinuities of $v_x$,  
we solve a scalar field $S$ governed by the advection equation 
$\partial S/\partial t + \bm{v} \cdot \bm{\nabla}S=0$.
The initial profile of $S$ is given by 
\begin{equation}
    S(y) = \frac{1}{2}
   \left\{
    \tanh\left( \frac{y+1/2}{w_0} \right)
    - \tanh\left( \frac{y-1/2}{w_0} \right)
    \right\}.
\end{equation}
The following perturbation is introduced in the vertical velocity field,
\begin{eqnarray}
    v_y(x,y) &=&
    \delta v \sin (k x) \\
    &\times & \left\{\exp\left( -\frac{(y+1/2)^2}{w_1^2} \right)
   + \exp\left( -\frac{(y-1/2)^2}{w_1^2} \right)
   \right\} \nonumber,
\end{eqnarray}
where $\delta v=0.01$ is the perturbation amplitude, $w_1=0.2$ is a parameter showing a spatial extent of the $v_y$ perturbation 
around the initial discontinuities and $k$ is the wavenumber and set to $2\pi$.
No perturbations are added in other variables.

We consider three different Alfv\'en Mach numbers, $\machA=8$, 
2, and $0.5$.
The non-dimensional parameter $k\LH$ is set to 4. 
All cases are in the Hall-dominated regime, and 
the growth rates $\sigma$ are close to $\sigma_\mathrm{hall}$ (Equation (\ref{sigmahall})).

\subsubsection{Comparison of Different Implementations}

\begin{figure}[htpb]
    \centering
    \includegraphics[width=8.5cm]{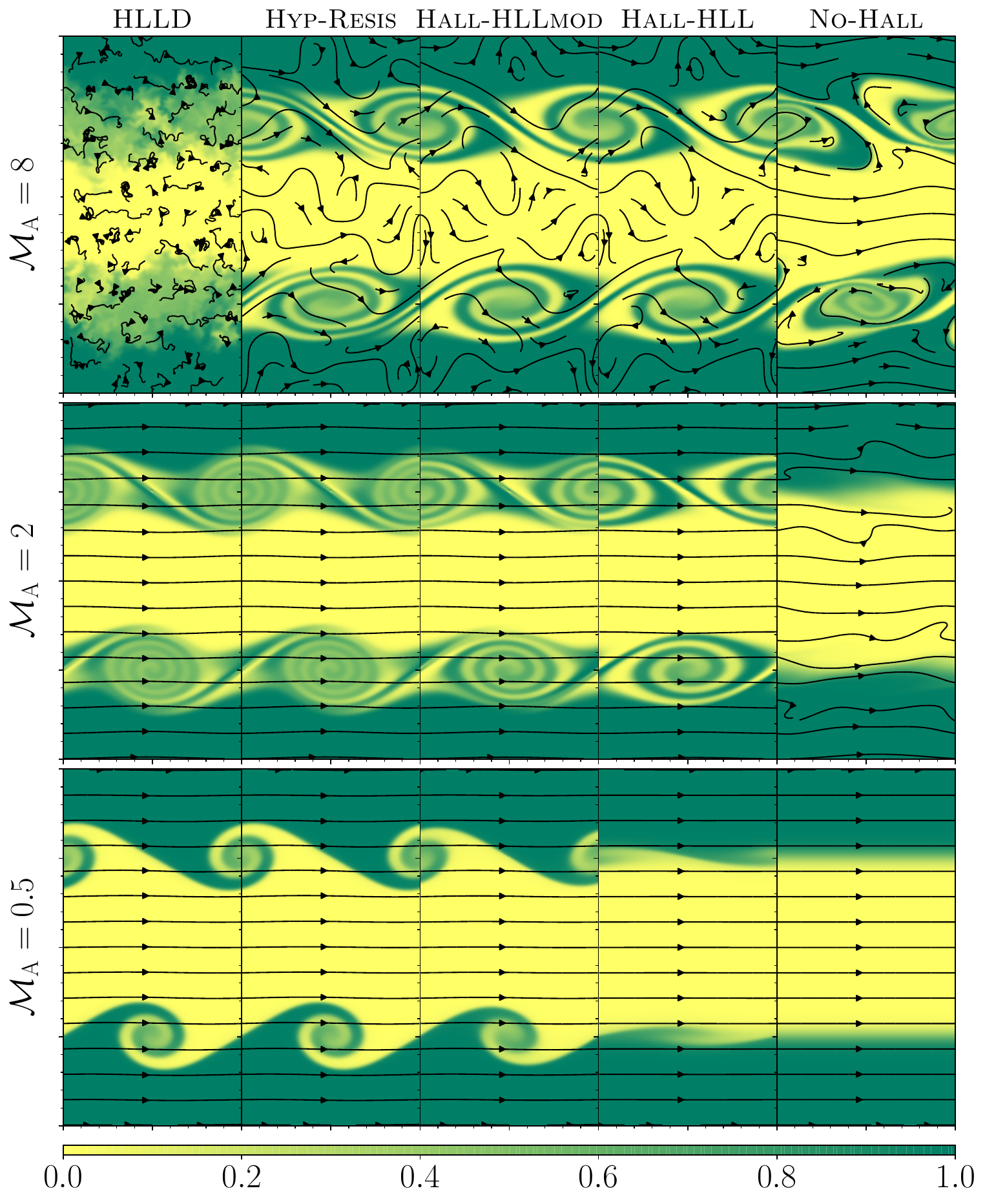}
    \caption{
    Snapshots of the scalar field $S$ at $t=4\machA^{-1}$ for 
    $\machA=0.8$ (top row), $\machA=2$ (middle row), and $\machA=0.5$ (bottom row).
    For each $\machA$, the results of 
    {\sc HLLD}, 
    {\sc Hyp-Resis} with $C_\mathrm{hyp}=0.05$, 
    {\sc Hall-HLLmod}, 
    {\sc Hall-HLL} are shown from left to right.
    In the rightmost column, the results without the Hall effect 
    are presented ({\sc No-Hall}).
    }
    \label{fig:kh_slice}
\end{figure}

\begin{figure}[htpb]
    \centering
    \includegraphics[width=8.5cm]{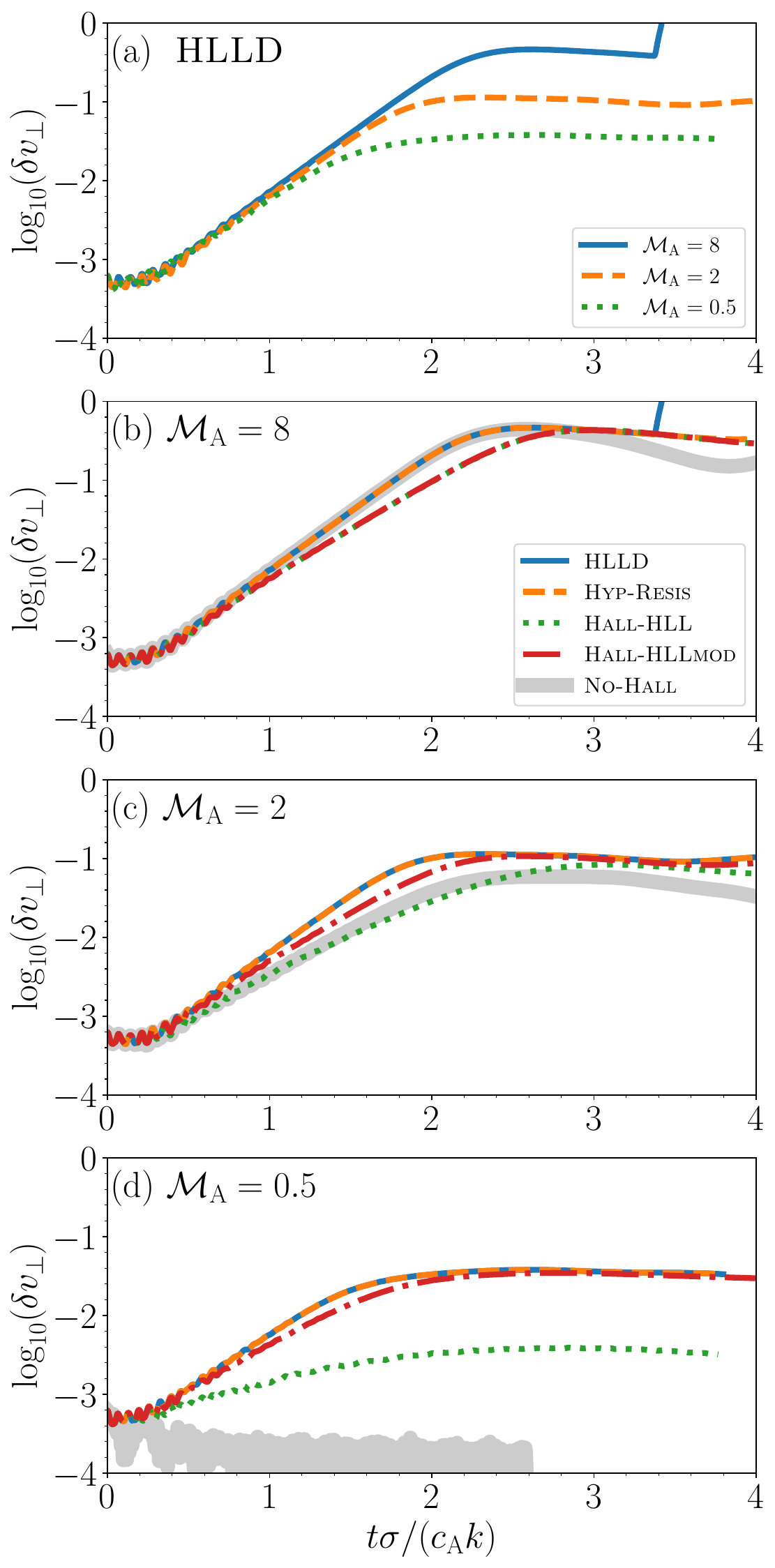}
    \caption{
    Time evolution of the vertical velocity perturbations 
    $\delta v_\perp = \sqrt{\langle v_y^2 + v_z^2\rangle}$.
    The horizontal axis represents the time normalized by $kc_\mathrm{A}\sigma^{-1}$ for $k\LH=4$ and $k=2\pi$.
    }
    \label{fig:dv_kh}
\end{figure}

Figures 
\ref{fig:kh_slice} 
and 
\ref{fig:dv_kh} show 
the $z=0$ slices of the scalar fields at $t=4\machA$ 
and 
the time evolution of $\delta v_\perp = \sqrt{\langle v_y^2 + v_z^2\rangle}$, respectively.

For references, the results without the Hall effect, which is labeled by {\sc No-Hall}, 
are shown in the rightmost column of Figure \ref{fig:kh_slice}.
Only the $\machA=0.5$ run without the Hall effect shows a stable result 
in Figure \ref{fig:kh_slice} (also see Figures \ref{fig:dv_kh}b, \ref{fig:dv_kh}c, and \ref{fig:dv_kh}d).
This is consistent with the results of the 
linear analysis by \citet{Chandrasekhar1961hhs..book.....C}, 

Figure \ref{fig:dv_kh}a shows that for {\sc HLLD},
the perpendicular velocity dispersions $\delta v_\perp$ 
with different $\machA$ grow following almost the same lines
in the range $\delta v_\perp \lesssim 10^{-2}$ when 
the time is normalized using 
the growth rate predicted from Equation (\ref{dispkh}).
Thus, the $\machA$ dependence of the growth rates is consistent with the results of 
the linear analysis, and the sub-Alf\'enic case ($\machA=0.5$) is destabilized by the Hall effect.
Note that the growth rates obtained from the simulations are about half of the predicted values.
This is probably because the settings of the simulations 
are not exactly the same as those of the linear analysis
presented above.
For instance, in the simulations, the smoothed profile of the shear flow is considered,
and the periodic boundary conditions are imposed
in all directions.

The results of {\sc HLLD} show the leftmost panels of 
Figure \ref{fig:kh_slice}.
Clear rolled-up vortices are seen only for $\machA=0.5$ 
and $2$, and the magnetic field lines are almost straight 
in the ($x$ - $y$) plane. 
The $\machA=8$ model suffers from the numerical instability due to the Hall effect.
This is because strong rolling-up motions bend the magnetic field lines significantly for 
$\machA=8$.
Onset of the numerical instability is observed as a sudden increase in $\delta v_\perp$ 
around $t \sigma/(c_\mathrm{A}k)\sim 3.4$ in Figure \ref{fig:dv_kh}a.

Next, the results for {\sc Hyp-Resis} with $C_\mathrm{hyp}=0.05$
are investigated.
For $\machA=2$ and 0.5, 
the $S$ maps are almost identical to those for {\sc HLLD}.
For $\machA=8$, 
the hyper-resistivity suppresses the numerical instability 
that occurs for {\sc HLLD}.
This feature is quantitatively evident in 
Figures \ref{fig:dv_kh}b-\ref{fig:dv_kh}d.
For all values of $\machA$,
the time evolution of $\delta v_\perp$ for {\cal Hyp-Resis} 
is almost identical to that for {\sc HLLD} except the numerical instability.
This indicates that hyper-resistivity
does not introduce significant dissipation.

In contrast, {\sc Hall-HLL} produces dissipative results, while
the amount of numerical dissipation depends on $\machA$.
For $\machA\ge 2$, the results of {\sc Hall-HLL} show clearly rolled-up vortices, although 
the number of rotations of the vortex sheet is lower than that for {\sc Hyp-Resis} (Figure \ref{fig:kh_slice}).
Figures \ref{fig:dv_kh}b and \ref{fig:dv_kh}c show that 
large numerical dissipation due to {\sc Hall-HLL} slows the growth of $\delta v_\perp$ compared to 
{\sc Hyp-Resis}.
For the sub-Alfvenic case ($\machA=0.5$), the development of vortices is strongly suppressed in 
Figure \ref{fig:kh_slice}.
The linear growth rate and saturation level of $\delta v_\perp$ are 
both significantly lower for {\sc Hall-HLL} than those for {\sc Hyp-Resis}.

These properties of {\sc Hall-HLL} are consistent with the results of the convergence test in 
the Hall-dominated regime ($k\LH=20$) shown in Section \ref{sec:linearwaves}.
The development of the KH instability for the super-Alfv\'enic cases 
is influenced by 
whistler waves, which are captured reasonably well by {\sc Hall-HLL}  
(see the fast-wave branch in Figure \ref{fig:convergence}).
However, for sub-Alfv\'enic cases, ion-cyclotron waves, 
which are strongly damped by {\sc Hall-HLL},
play an important role in the development of the KH instability.

In a similar way to Section \ref{sec:linearwaves},
{\sc Hall-HLLmod} significantly 
improves the dissipative results of {\sc Hall-HLL}, 
especially for smaller $\machA$.
However, {\sc Hall-HLLmod} produces more dissipative results than {\sc Hyp-Resis}.
We will publically release our implementation of the Hall effect shortly.

\section{Conclusions}

In this paper, we evaluate the performance of 
several numerical methods for
the Hall effect found in the literature, 
which are listed in Table \ref{tab:methods},
based on an extensive series of test calculations.
The Hall effect is implemented in \texttt{Athena++} \citep{Stone2020ApJS..249....4S} (see Section \ref{sec:implementation}).
Two types of implementations of the Hall effect are considered.
One is {\sc Hall-HLL}, where the phase speed of whistler waves 
is taken into account to compute the signal speeds in the HLL 
numerical fluxes \citep{LesurKunzFromang2014}.
The modified version of {\sc Hall-HLL} ({\sc Hall-HLLmod}) proposed by \citet{Marchand2019AA...631A..66M} is also tested.
In {\sc Hall-HLLmod}, the hydrodynamical variables 
(the density, momentum, and total energy) are updated by 
using the original HLL numerical fluxes, whereas 
the {\sc Hall-HLL} numerical fluxes are used to update the magnetic field.
The other implementation ({\sc Hyp-Resis}) 
introduces a fourth order 
hyper-resistivity into the induction equation.

An appropriate hyper-resistivity coefficient 
($C_\mathrm{hyp}\sim 0.05$) is
determined to ensure both numerical 
stability and high accuracy based on numerical experimants 
and the von Neumann stability analysis 
(Sections \ref{sec:Chypmin} and \ref{sec:denshear}, 
Appendix \ref{app:vonNeumann}).

The difference in the performance of the methods is 
clearly observed in the convergence test of linear waves (see
Section \ref{sec:linearwaves}).
In the Hall-dominated regime, 
{\sc Hyp-Resis} exhibits second-order convergence for all 
types of Hall-MHD linear waves, and numerical dissipation 
caused by the hyper-resistivity term does not significantly affect the 
accuracy of the solutions.
By contrast, {\sc Hall-HLL} shows 
significantly slower convergence than second-order 
due to numerical dissipation in the ion-cyclotron wave and 
sound wave, whereas it exhibits second-order convergence for 
the whistler wave.
This occurs because the numerical dissipation in {\sc Hall-HLL}
is determined by the fastest phase speed among the 
linear waves, which is the whistler-wave phase speed.
This behavior is also seen in the Kelvin-Helmholtz 
instability test (Section \ref{sec:KH}).
{\sc Hall-HLLmod} significantly reduces numerical dissipation 
compared to {\sc Hall-HLL}, but produces more diffusive results than {\sc Hyp-Resis}.

Section \ref{sec:uniformgrid} demonstrated that 
{\sc Hyp-Resis} with $C_\mathrm{hyp}\sim 0.05$ 
is a suitable choice also in terms of computational performance.
The computational cost of calculating the hyper-resistivity term is negligible 
compared to the total cost.
{\sc Hyp-Resis} is slightly slower than {\sc Hall-HLLmod} for the same resolution, but it can achieve the same accuracy with considerably fewer grid points. In other words, it can achieve a better accuracy with the same computational cost.

In summary, hyper resistivity with an appropriate coefficient ensures both numerical stability and high accuracy, 
making it the optimal choice for simulating 
phenomena involving the Hall effect.

\begin{acknowledgments}
Numerical computations were carried out on 
Cray XC50 and XD2000 at the CfCA of the   National Astronomical Observatory of Japan.
This work was supported in part by the Ministry of Education, Culture, Sports, Science and Technology (MEXT), Grants-in-Aid for Scientific Research, JP21H00056 (K.I.), JP16H05998, JP21H04487, JP22KK0043 (K.T. and K.I.). 
This research was also supported by MEXT as ``Program for Promoting Researches on the Supercomputer Fugaku” (Toward a unified view of the universe: from large scale structures to planets, JPMXP1020200109) 
and ``Structure and Evolution of the Universe Unraveled by Fusion of Simulation and AI" (JPMXP1020230406).

\end{acknowledgments}

{\software{ Athena++ \citep{Stone2020ApJS..249....4S}, numpy \citep{Numpy}, Matplotlib \citep{Matplotlib}}}

%



 

\appendix

\section{von Neumann Analysis for Hall-MHD}\label{app:vonNeumann}

We present the results of the von Neumann analysis for Hall MHD.
For whistler waves whose wavenumbers satisfying $\etaH k \gg c_\mathrm{A}$,
the ideal term $\bm{\nabla}\times (\bm{v}\times \bm{B})$ is much smaller than 
the Hall term in the induction equation. 
In addition, the gas is nearly static because 
the velocity perturbations $\delta v$ are 
much smaller than $\delta B/\sqrt{4\pi\rho_0}$.
Thus, we consider the following induction equation,
\begin{equation}
    \frac{\partial \bm{B}}{\partial t}
    = - \bm{\nabla} \times \left(
     \etaH \frac{(\bm{\nabla}\times \bm{B}) \times\bm{B}}{|\bm{B}|}
    \right).
    \label{induc_app}
\end{equation}

The computational volume is divided into cells with a size of
$\Delta x \Delta y \Delta z$.
We assume uniform grids for simplicity.
The cell centers are defined at $(i\Delta x,j\Delta y,k\Delta z)$, and 
the variable $Q$ is denoted as $Q_{i,j,k}$.
The magnetic fields are defined at the cell surfaces; 
$(B_x)_{i-1/2,j,k}$,
$(B_y)_{i,j-1/2,k}$, and $(B_z)_{i,j,k-1/2}$.

As the unperturbed state, we consider a static gas 
with constant density and pressure and 
a uniform magnetic field of 
$\bm{B}_0=(B_{x0},B_{y0},B_{z0})$.
A plane wave perturbation with a wavenumber vector 
of $\bm{\kappa} = (\kappa_x,\kappa_y,\kappa_z)$ is considered.
The magnetic field components are given by 
\begin{eqnarray}
(B_x)_{i-\half,j,k} &=& B_{x0} + \delta B_x e^{
I(\alpha_x (i-\half) + \alpha_y j + \alpha_z k)},\nonumber \\
(B_y)_{i,j-\half,k} &=& B_{y0} + \delta B_y e^{
I(\alpha_x i + \alpha_y (j-\half) + \alpha_z k)},\\
(B_z)_{i,j-\half,k} &=& B_{z0} + \delta B_z e^{
I(\alpha_x i + \alpha_y (j-\half) + \alpha_z k)},\nonumber 
\label{Bperturb}
\end{eqnarray}
where $\alpha_x = \kappa_x \Delta x$, $\alpha_y = \kappa_y \Delta y$, $\alpha_z=\kappa_z\Delta z$, and 
$I$ is the imaginary unit.
Substituting Equation (\ref{Bperturb}) into the discretized form of Equation 
(\ref{induc_app}),
one obtains 
\begin{equation}
\frac{\partial \delta \bm{B}}{\partial t}
= 
{\cal R}\delta \bm{B}
\end{equation}
where $\delta \bm{B}=(\delta B_x,\delta B_y,\delta B_z)$,
\begin{equation}
{\cal R} = 
2\etaH C 
\left(
\begin{array}{ccc}
0 & -\sin(\alpha_z/2)/\Delta z & {\sin(\alpha_y/2)}/{\Delta y}\\
{\sin(\alpha_z/2)}/{\Delta z} & 0 & -{\sin(\alpha_x/2)}/{\Delta x} \\
-{\sin(\alpha_y/2)}/{\Delta y} & {\sin(\alpha_x/2)}/{\Delta x} & 0 \\
\end{array}
\right)
\end{equation}
and
\begin{equation}
C = 2\prod_{l=x,y,z}\cos\left(\frac{\alpha_l}{2}\right)
\sum_{l=x,y,z}
\left[
\frac{B_{l0}\tan(\alpha_l/2)}{B_0\Delta l}
\right].
\end{equation}
$C$ is reduced to $\bm{k}\cdot\bm{B}/B_0$ in the small wavenumber limit ($\alpha_l\ll1$).

With third-order time integrators, one obtains 
$\delta \bm{B}^{n+1}={\cal Q}\delta \bm{B}^n$, where 
${\cal Q} = {\cal I}+{\cal R}\Delta t+ {\cal R}^2 \Delta t^2/2 + {\cal R}^3\Delta 
t^3/6$.
The discrete form is stable if the absolute values of 
the eigenvalues $\lambda$ of ${\cal Q}$ is less than unity, or 
\begin{equation}
  |\lambda | = \sqrt{\left(1-\frac{\xi_\mathrm{H}^2}{2}\right)^2+
    \left(\xi_\mathrm{H}-\frac{\xi_\mathrm{H}^3}{6}\right)^2} < 1,
    \label{criterion_rk3}
\end{equation}
where 
\begin{equation}
    \xi_\mathrm{H} = \etaH\Delta t C\sqrt{ \sum_{l=x,y,z} 
    \left(\frac{2\sin(\alpha_l/2)}{\Delta l} \right)^2}.
\end{equation}
For small wavenumber limits, Equation (\ref{criterion_rk3}) reduces to that derived by 
\citet[][see their Appendix B]{Kunz2013MNRAS.434.2295K}.

Equation (\ref{criterion_rk3}) is reduced to
$\xi_\mathrm{H} <\sqrt{3}$.
An conservative criterion for ensuring the stability of any linear wave is 
\begin{equation}
    \frac{4\sqrt{3}\etaH \Delta t}{\min(\Delta x^2, \Delta y^2, \Delta z^2)} < \sqrt{3},
\end{equation}
where we use the fact that $C<2/\min(\Delta x,\Delta y,\Delta z)$ and 
$\sin(\alpha_l/2)<1$.

For second-order time integrators
where 
${\cal Q} = {\cal I}+{\cal R}\Delta t+ {\cal R}^2 \Delta t^2/2$,
$|\lambda|$ becomes $\sqrt{1+\xi_\mathrm{H}^2/4}$, 
indicating that the discretization form is unconditionally unstable \citep{Falle2003MNRAS.344.1210F,Kunz2013MNRAS.434.2295K}.


\bibliography{ms_submit}{}
\bibliographystyle{aasjournal}



\end{document}